\documentclass[prc,showpacs,preprintnumbers]{revtex4-2}
\usepackage{amsmath}
\usepackage{amssymb}
\usepackage{graphicx}
\usepackage{bm}
\usepackage{comment}
\usepackage{mathrsfs}
\usepackage{subeqnarray}
\usepackage{hyperref}
\usepackage{color}
\usepackage[normalem]{ulem} 
\usepackage{fancyhdr}
\usepackage{datetime}
\makeatletter
\newcommand{\doitext}{DOI:\ }
\newcommand{\doiurl}{https://doi.org/}
\renewcommand*{\doi}{%
  \begingroup 
  \lccode`\~=`\#\relax 
  \lowercase{\def~{\#}}%
  \lccode`\~=`\_\relax
  \lowercase{\def~{\_}}%
  \lccode`\~=`\<\relax 
  \lowercase{\def~{\textless}}%
  \lccode`\~=`\>\relax 
  \lowercase{\def~{\textgreater}}%
  \lccode`\~=0\relax 
  \catcode`\#=\active 
  \catcode`\_=\active 
  \catcode`\<=\active 
  \catcode`\>=\active 
  \@doi
}
\def\@doi#1{%
  \let\#\relax
  \let\_\relax
  \let\textless\relax 
  \let\textgreater\relax 
  \edef\x{\toks0={{#1}}}%
  \x
  \edef\#{\@percentchar23}%
  \edef\_{_}%
  \edef\textless{\@percentchar3C}
  \edef\textgreater{\@percentchar3E}
  \edef\x{\toks2={\noexpand\href{\doiurl#1}}}%
  \x
  \edef\x{\endgroup\doitext\the\toks2 \the\toks0}%
  \x
}
\makeatother
\numberwithin{equation}{section}
\usepackage{ifpdf} 
\begin{document}
\ifpdf
 \def\anglefig{0}
 \def\scalefig{1}
\else
 \def\anglefig{-90}
 \def\scalefig{0.75}
\fi
\title{Exact expressions for the number of levels in single-\textit{j} orbits 
for three, four and five fermions}
\author{Michel Poirier}
\email{michel.poirier@cea.fr}
\affiliation{CEA - Paris-Saclay University, Laboratory ``Interactions, Dynamics, 
and Lasers'', CE Saclay, F-91191 Gif-sur-Yvette, France}
\author{Jean-Christophe Pain}
\email{jean-christophe.pain@cea.fr}
\affiliation{CEA, DAM, DIF, F-91297 Arpajon, France}
\affiliation{Universit\'e Paris-Saclay, CEA, Laboratoire Mati\`ere en Conditions Extr\^emes, F-91680 Bruy\`eres-le-Ch\^atel, France}
\date{\today}
\newcommand{\sixj}[6]{\left\{\begin{array}{ccc}#1 & #2 & #3 \\ #4 & #5 & #6 \end{array}\right\}}
\newcommand{\ninej}[9]{\left\{\begin{array}{ccc}#1 & #2 & #3 \\ #4 & #5 & #6 \\ #7 & #8 & #9 \end{array}\right\}}


\begin{abstract}
We propose closed-form expressions of the distributions of magnetic quantum number $M$ and total angular momentum $J$ for three and four fermions in single-$j$ orbits. The latter formulas consist of polynomials with coefficients satisfying congruence properties. Such results, derived using doubly-recursive relations over $j$ and the number of fermions, enable us to deduce 
explicit expressions for the total number of levels in the case of three-, four- and five-fermion systems. We present applications of these formulas, such as sum rules for six-$j$ and nine-$j$ symbols, obtained from the connection with fractional-parentage coefficients, an alternative proof of the Ginocchio-Haxton relation or cancellation properties of the number of levels with a given angular momentum. 
\par\medskip
\noindent 
\doi{10.1103/PhysRevC.104.064324}
\end{abstract}

\maketitle

\section{Introduction}

Determining the allowed total angular momenta $J$ to which the individual 
half-integer spins $j$ of $N$ identical particles 
may couple is of primary importance in nuclear physics. Some values of $J$ are forbidden by 
the Pauli exclusion principle, others occur more than once. Although that problem was investigated by many authors over the years, and despite the variety of approaches (number theory, recurrence relations, generating functions, etc.), exact analytical expressions for the number of states $P(M)$ 
with a given projection $M$ on the quantization axis, 
the number of levels $Q(J)$ with spin $J$ 
or the total number of levels $Q_{\text{tot}}$ 
in a configuration are not known, except in very simple cases.

Zhao and Arima have shown that there are simple structures in $Q(J)$ for $j^3$ or $j^4$, and found empirical formulas \cite{Zhao2003}. In 2005, the same authors \cite{Zhao2005} showed that $Q(J)$ could be enumerated by the reduction from $SU(N+1)$ to $SO(3)$ and obtained analytical expressions of $Q(J)$ for four particles. 
The same year, Talmi derived a recursion formula for $Q(J)$ \cite{Talmi2005}. 
The latter quantity for $j^N$ is expressed in terms of $Q(J)$ for $(j-1)^N$, $(j-1)^{N-1}$ and $(j-1)^{N-2}$. In the same work, Talmi also proved some interesting results found empirically by Zhao and Arima \cite{Zhao2003}. Zhang \emph{et al} extended Talmi's recursion relation to boson systems and proved empirical formulas for five bosons. They also obtained the number of states with given spin for three and four bosons by using sum rules of six-$j$ and nine-$j$ symbols \cite{Zhang2008,Pain2011}. Five years later, Jiang \emph{et al} derived the analytical formulas for $Q(J)$ for three fermions in a single-$j$ shell and three bosons with spin $\ell$, by using a reduction rule from the $U(4)$ to the $O(3)$ group chain, $U(4) \supset Sp(4) \supset O(3)$ \cite{Hamermesh1962}, for $\tilde{N}$ virtual bosons which follow the $U(4)$ symmetry (i.e., spin 3/2) \cite{Jiang2013}. One has $\tilde{N}=2j-2$ for fermions and $\tilde{N}=2\ell$ for bosons. The authors were able to obtain analytical formulas of three bosons and fermions in a unified form and on a unified footing. 
Let us consider a system of $N$ identical fermions in a single 
$j$ (which is half-integer) shell of degeneracy $g=2j+1$, $m_i$ being the angular momentum 
projection of electron state $i$ ($m_1=-j, m_2=-j+1, m_3=-j+2, \cdots, m_{g-1}=j-1, m_g=j$). 
The maximum total angular momentum is
\begin{equation}J_{\text{max}}=(2j+1-N)N/2\end{equation}
and the minimum angular momentum $J_{\text{min}}$ is 0 if $N$ is even and $1/2$ if $N$ is odd.
The distribution $P(M)$ represents the number of $N$-fermion states having the total projection 
(or magnetic quantum number) $M$. 
The number $Q(J)$ of levels with angular momentum $J$ in a configuration can be obtained from the 
distribution $P(M)$ of the $M$ values by means of the relations \cite{Bethe1936,*Landau1977}
\begin{subequations}\begin{gather}\label{eq:QvsP}
 Q(J)=P\left(J\right)-P\left(J+1\right)\text{\quad if }J\le J_{\text{max}}-1,\\
 Q(J_{\text{max}})=P(J_{\text{max}}).\end{gather}
\end{subequations}
In the following we use the notation $P(M;j,N)$ instead of $P(M)$ everytime it is necessary 
to specify the angular momentum of the shell and the number of fermions.

The fundamental relation used in the present paper to get the number of states 
$P(M;j,N)$ of $N$ fermions with spin $j$ and total magnetic quantum number $M$ 
has been derived by Talmi [Eq.~(1) in Ref.~\cite{Talmi2005}]
\begin{equation}\label{eq:Talmi}
 P(M;j,N)=P(M;j-1,N)+P(M-j;j-1,N-1)+P(M+j;j-1,N-1)+P(M;j-1,N-2).
\end{equation}
A short alternative derivation is presented in Appendix \ref{sec:Talmirec}. 
From the above relation (\ref{eq:QvsP}), one also gets easily the total number of levels
\begin{equation}\label{eq:totlev}
\sum_{J=J_\text{min}}^{J_\text{max}}Q(J;j,N)=P\left(J_{\text{min}};j,N\right)
\end{equation}
where $J_\text{min}=0$ (resp. 1/2) for $N$ even (resp. odd).
A simple expression for the total number of levels for $j^3$ was found using coefficients of fractional parentage \cite{Pain2019}. In the case of four fermions, no explicit formula could be obtained with the latter technique, only a triple summation involving nine-$j$ coefficients, or equivalently products of two six-$j$ symbols multiplied by Dunlap-Judd coefficients \cite{Dunlap1975}. 

In the present work, using the recurrence relation (\ref{eq:Talmi}), we derive  
explicit expressions for $P(M;j,3)$, $Q(J;j,3)$ (Section \ref{sec:PMj3}), $P(M;j,4)$, 
and $Q(J;j,4)$ (Section \ref{sec:PMj4}), as well as for the total number of 
$J$-levels in the case of five fermions (Section \ref{sec:Qtotj5}). This leads us 
to deduce exact formulas for $Q_{\text{tot}}\left(j^3\right)$ (i.e., an alternative 
derivation much simpler than the one previously published and relying on the use of 
fractional parentage coefficients \cite{Pain2019}), for $Q_{\text{tot}}(j^4)$ and 
for $Q_{\text{tot}}(j^5)$. To our knowledge, no expressions of the two latter 
formulas were published elsewhere. 
The algebraic forms of $Q(J;j,3)$ and $Q(J;j,4)$ are also likely to yield to sum 
rules for six-$j$ symbols (Section \ref{sec:sum}). We also provide some additional 
results, such as an alternative derivation of the Ginocchio-Haxton relation 
(Section \ref{sec:sum}), cancellation properties and particular values of the 
number of levels with a given angular momentum (Section \ref{sec:part}).

\section{Three-fermion systems}
\label{sec:PMj3}
\subsection{Total number of levels}\label{sec:Qtotj3}
The total number of levels will be derived from Eq.~(\ref{eq:totlev}). For three 
particles, the relation (\ref{eq:Talmi}) is written as
\begin{equation}\label{eq:PN3s4t}
P\left(\frac12;j,3\right)=P\left(\frac12;j-1;3\right)+P\left(\frac12-j;j-1,2\right)
+P\left(\frac12+j;j-1,2\right)+P\left(\frac12;j-1,1\right).
\end{equation}
This provides us with a recurrence relation on $j$ for $P(1/2;j,3)$, which is 
initialized by the value $P(1/2;3/2,3)$. Using the relation easily obtained by 
considering the coupling of two momenta
\begin{equation}\label{eq:PMj2}
P(M;j,2)=\left\lfloor\frac{2j+1-|M|}{2}\right\rfloor
\end{equation}
where $\lfloor x\rfloor$ is the integer part of $x$, we get immediately, for $j$ 
half-integer,
\begin{subequations}\begin{gather}
P\left(\frac12-j;j-1,2\right)=\left\lfloor\frac{j}{2}-\frac14\right\rfloor\\
P\left(\frac12+j;j-1,2\right)=\left\lfloor\frac{j}{2}-\frac34\right\rfloor
\end{gather}\end{subequations}
and a rapid inspection of the cases $j=2n+1/2, j=2n+3/2$ shows that, since
$P(1/2;j-1,1)=1$ for $j\ge3/2$, one has
\begin{equation}
P\left(\frac12;j,3\right)=P\left(\frac12;j-1,3\right)+j-\frac12
\end{equation}
for $j\ge3/2$. Since the coupling of three angular momenta $j=1/2$ is not possible (Pauli exclusion principle), we have $P(1/2;1/2,3)=0$ and therefore
\begin{equation}\label{qtotj3}
Q_{\mathrm{tot}}\left(j^3\right)=P\left(\frac12;j,3\right)
 =\sum_{i=1/2}^j(i-1/2)=\sum_{t=0}^{j-1/2}t=\frac12\left(j^2-\frac14\right)
\end{equation}
in agreement with the formula (36) of Ref.~\cite{Pain2019}.

\subsection{Determination of the \textit{M} distribution for three fermions}
\label{eq:PMj3}
\subsubsection{Case \textit{M} greater than \textit{j}}
We first determine $P(j+q;j,3)$ with $q$ positive integer ($q=1,2\dots J_\text{max}-j$). Using Talmi's formula and the explicit value (\ref{eq:PMj2}) 
one gets, after $p$ iterations,
\begin{subequations}\begin{align}
 P(j+q;j,3) &= P(j+q;j-1,3)+P(q;j-1,2)\\
 &= P(j+q;j-1,3)+\left\lfloor j-\frac{q+1}{2}\right\rfloor\\
 &= P(j+q;j-2,3)+\left\lfloor j-\frac{q+1}{2}-\frac32\right\rfloor
 +\left\lfloor j-\frac{q+1}{2}\right\rfloor\\
 &\vdots\nonumber\\
 &= P(j+q;j-p,3)+\sum_{t=0}^{p-1}\left\lfloor j-\frac{q+1}{2}-\frac{3t}{2}\right\rfloor,
 \label{eq:PjpqjN3rec}
\end{align}
\end{subequations}
where we have used the property $P(2j+q-t;j-t,2)=0$ and $P(j+q-t;j-t,1)=0$ 
valid for $q>0$, and $0\le t\le p-1$. 
We choose $p$ such that $P(j+q;j-p,3)$ vanishes 
while $P(j+q;j-p+1,3)$ does not. This yields the conditions
\begin{equation}\label{eq:condp}
1\le j-\frac{q}{2}-\frac{3p}{2}+1,\quad 
j-\frac{q}{2}-\frac{3p}{2}-\frac12 < 1,
\end{equation}
which amount to 
\begin{equation}\label{eq:p_fnjmq_N3}
p=\left\lfloor\frac{2j-q}{3}\right\rfloor.
\end{equation}
Since $q$ can be even or odd, for $j$ half-integer $j-q/2$ is either integer or 
half-integer. When evaluating $\lfloor(2j-q)/3\rfloor$ six cases must be 
considered. One obtains for the value of the maximum index $p$
\begin{equation}\label{eq:pmaxj3}
 p = \begin{cases}
  2n & \quad\text{if }j-\frac{q}{2}=3n, 3n+1/2,\text{ or }3n+1\\
  2n+1 & \quad\text{if }j-\frac{q}{2}=3n+3/2,3n+2,\text{ or }3n+5/2.\\
 \end{cases}
\end{equation}
In the computation of the sum (\ref{eq:PjpqjN3rec}) with that value of $p$, we 
note that $P(j+q;j-p,3)$ vanishes because of the conditions (\ref{eq:condp}). 
We distinguish six cases, according to the maximum index (\ref{eq:pmaxj3}). 
For instance if $j-q/2=3n$ the sum is, after reordering odd and even $t$ values,
\begin{subequations}\begin{align}
P(j+q;j,3) &= \lfloor1\rfloor+\lfloor5/2\rfloor+\lfloor4\rfloor+\cdots
 \lfloor 3n-2\rfloor+\lfloor 3n-1/2\rfloor\\
 &=1+4+\cdots+(3n-2)+2+5+\cdots+(3n-1)\\
 &= \sum_{t=1}^n (3t-2)+\sum_{t=1}^n (3t-1)=3n^2
  =\frac13\left(j-\frac{q}{2}\right)^2.
\end{align} \end{subequations}

\begin{table}[htb]
\centering\renewcommand*{\arraystretch}{1.75}
\begin{tabular}{ccccccc}
\hline\hline
$j-q/2$ & $3n$ & $3n+1/2$ & $3n+1$ & $3n+3/2$ & $3n+2$ & $3n+5/2$ \\
\hline
First term  & $\lfloor1\rfloor$ & $\lfloor3/2\rfloor$ & $\lfloor2\rfloor$ &
 $\lfloor1\rfloor$ & $\lfloor3/2\rfloor$ & $\lfloor2\rfloor$\\
Last term  & $\lfloor3n-1/2\rfloor$ & $\lfloor3n\rfloor$ & $\lfloor3n+1/2\rfloor$
 & $\lfloor3n+1\rfloor$ & $\lfloor3n+3/2\rfloor$ & $\lfloor3n+2\rfloor$\\
Sum & $3n^2$ & $n(3n+1)$ & $n(3n+2)$ & $3n(n+1)+1$ & $(n+1)(3n+1)$ & $(n+1)(3n+2)$\\
\hline\hline
\end{tabular}
\caption{Various cases for the computation of $P(j+q;j,3)$}\label{tab:calcPjq3}
\end{table}

The six cases are summed up in Table \ref{tab:calcPjq3}.
Expressing $n$ versus $j-q/2$, we obtain the desired formula
\begin{subequations}\begin{align}\label{eq:PjpqjN3}
 P(j+q;j,3) &= \frac13\left(j-\frac{q}{2}\right)^2+\alpha(2j-q)\\
  \text{with }\alpha(2j-q)&=\left(0,-\frac{1}{12},-\frac{1}{3},\frac{1}{4},
  -\frac{1}{3},-\frac{1}{12}\right) \text{if }2j-q\bmod6=(0,1,2,3,4,5)
  \text{ respectively.}
\end{align}\end{subequations} 
For instance one can check for $q=1$
\begin{equation}\label{eq:Pj1jN3}
 P(j+1;j,3) = \begin{cases}
  \frac13\left(j-\frac{1}{2}\right)^2 & \text{if }j-1/2=3n\\
  \frac13\left(j-\frac{3}{2}\right)\left(j+\frac{1}{2}\right)=
  \frac13\left(j-\frac{1}{2}\right)^2 -\frac13 & 
  \text{if }j-1/2=3n+1 \text{ or }j-1/2=3n+2.
 \end{cases}
\end{equation}
The formula (\ref{eq:PjpqjN3}) does not assume that $j$ is half-integer. 
Instead of (\ref{eq:Pj1jN3}), we would have, for integer $j$,
\begin{equation}
 P(j+1;j,3) = \begin{cases}
  \frac13\left(j-\frac{1}{2}\right)^2-\frac{1}{12}= \frac13(j-1)j
  & \text{if }j=3n+1 \text{ or if }j=3n\\
  \frac13\left(j-\frac{1}{2}\right)^2+\frac14=\frac13(j^2-j+1) 
  & \text{if }j=3n+2.
 \end{cases}
\end{equation}

\subsubsection{Case \textit{M} less than or equal to \textit{j}}
In this section one assumes $j$ half-integer such that $j\ge3/2$. From the basic 
relation (\ref{eq:Talmi}), one writes
\begin{subequations}\label{eq:TalmiN3}\begin{gather}
 P(j-q;j,3)=P(j-q;j-1,3)+S(j,q)\label{eq:recPjmq}\\
 \text{where }S(j,q)=P(2j-q;j-1,2)+P(q;j-1,2)+P(j-q;j-1,1).\label{eq:defSjq}
\end{gather}\end{subequations}
The quantity $S(j,q)$ is easily transformed using the value (\ref{eq:PMj2}) and the 
fact that $P(j-q;j-1,1)=1$ if $q>0$. Using this definition one easily checks that the 
terms in $S(j,0)$ take the values 0, $j-1/2$, 0 respectively, so that $S(j,0)=j-1/2$. 
If $q>0$ the identity (\ref{eq:PMj2}) provides the result 
\begin{equation}
 S(j,q)=\lfloor(q-1)/2\rfloor+\lfloor j-(q+1)/2\rfloor+1
\end{equation}
and considering the cases $q$ even or odd one easily verifies that, for $j\ge3/2$,
\begin{equation} S(j,q)=j-\frac12\end{equation}
which is also valid if $q=0$. The formula (\ref{eq:TalmiN3}) leads to a recurrence 
relation
\begin{subequations}\begin{align}
 P(j-q;j,3)&= P(j-q;j-1,3)+j-\frac12\\
  &= P(j-q;j-2,3)+j-\frac32+j-\frac12\\
 &\vdots\nonumber\\
  &= P(j-q;j-q-1,3)+\sum_{s=0}^q\left(j-\frac12-s\right)\\
  &= P(j-q;j-q-1,3)+(q+1)\left(j-\frac{q+1}{2}\right).
\end{align}\end{subequations}
The initial value $ P(j-q;j-q-1,3)$ is derived from the expression (\ref{eq:Pj1jN3}). 
One finds for $0\le q\le j-1/2$
\begin{equation}\label{eq:PjmqjN3}
P(j-q;j,3)=(q+1)\left(j-\frac{q+1}{2}\right)+\begin{cases}
 \frac13\left(j-q-\frac32\right)^2 & \text{ if } j-q=3n+\frac32\\
 \frac13\left(j-q-\frac52\right)\left(j-q-\frac12\right)
 &\text{ if } j-q=3n+\frac12 \text{ or }3n+\frac52\end{cases}
\end{equation}
or after simplification
\begin{subequations}\begin{align}\label{eq:PjmqjN3s}
P(j-q;j,3)&=\frac13\left(j+\frac{q}{2}\right)^2-\frac{q^2}{4}+\beta(j-q-1/2)\\
 \text{with }\beta(n) &=\left(-\frac{1}{12},\frac14,-\frac{1}{12}\right),
 \text{ if }n\bmod3=(0,1,2)\text{ respectively.}
\end{align}\end{subequations}
For instance, one obtains in the $q=0$ case
\begin{equation}\label{eq:PjjN3}
P(j;j,3)=\frac{j^2}3\begin{cases} +\frac14 &\text{ if } j=3n+3/2\\
 -\frac{1}{12}=\frac13\left(j^2-\frac14\right)
 &\text{ if } j=3n+1/2 \text{ or }3n+5/2.\end{cases}
\end{equation}
Such formulas can be generalized for $j$ integer but 
the resulting expressions will be different.
The formula (\ref{eq:PjmqjN3}) was established for $q\ge0$. One can check that it 
remains true for $q=-1$. Assuming (\ref{eq:PjmqjN3}) is valid for $q=-1$ we get
a piece-wise expression which is identical to (\ref{eq:Pj1jN3}).
It is worth mentioning that the relation (\ref{eq:PjmqjN3}) applies in particular 
for $j-q=1/2, 3/2,\dots n+1/2$. A series of examples is provided in Appendix 
\ref{sec:PlowMj3}. Finally if $j-q=n+1/2$, with $0\le n\le j-1/2$ one has
\begin{equation}
 P(n+1/2;j,3)=P(n+1/2;n-1/2,3)+\frac12\left(j^2-\left(n-\frac12\right)^2\right).
\end{equation}

\subsubsection{General case}
The formulas (\ref{eq:PjpqjN3}), (\ref{eq:PjmqjN3s}) can be gathered in a single 
equation, valid for any integer $q$. Using the Heaviside function $H(q)=1$ if 
$q\ge0$, 0 otherwise, one has
\begin{equation}
 P(j-q;j,3)=\frac13\left(j+\frac{q}{2}\right)^2-H(q)\frac{q^2}{4}
 +H(q)\beta(j-q-1/2)+(1-H(q))\alpha(2j+q).
\end{equation}
Considering the various values of $2j+q\bmod6$ and $q\bmod2$, one can then 
easily check that $\gamma=\alpha(2j+q)-\beta(j-q-1/2)$ is indeed a function of 
$q$ and equal to $-(q\bmod2)/4$. The above equation transforms into
\begin{subequations}\begin{align}\label{eq:PMjN3gen}
 P(j-q;j,3)&=\frac13\left(j+\frac{q}{2}\right)^2+\alpha(2j+q)
  -H(q)\left[\frac{q^2}{4}+\gamma(q)\right]\\
 \text{ with }\gamma(q)&=\left(0,-\frac14\right)\text{ for }q\bmod2=(0,1)
\end{align}\end{subequations}
if $-2j+3\le q\le j-1/2$, and $\alpha$ defined above (\ref{eq:PjpqjN3}).

\subsection{Distribution of the total angular momentum}
Using the fundamental relation (\ref{eq:QvsP}), the expression (\ref{eq:PjpqjN3}) 
allows us to derive the distribution of the total momentum $J$. The evaluation of 
$P(j+q;j,3)-P(j+q+1;j,3)$ provides
\begin{align}\label{eq:QjpqjN3}\nonumber
 Q(j+q;j,3) &= \frac{2j-q}{6}+q_{3p}\\\text{ with } q_{3p}&=
\left(0,-\frac{1}{6},-\frac{1}{3},\frac{1}{2},-\frac{2}{3},\frac{1}{6}\right)
 \text{ for }2j-q\bmod6=(0,1,2,3,4,5)\text{ respectively.}
\end{align}
For instance one has $Q(j;j,3)=j/3+(-1/6,1/2,1/6)$ if $j-1/2\bmod3=0,1,2$
respectively, i.e., $Q(j;,j,3)=\lfloor(2j+1)/6\rfloor$. One also verifies that 
$Q(j+1;j,3)=(2j-1)/6+(0,-1/3,-2/3)$ for $j-1/2\bmod3=0,1,2$ respectively, i.e., 
$Q(j+1,j,3)=\lfloor(2j-1)/6\rfloor$. Similarly, from (\ref{eq:PjmqjN3}), the 
evaluation of $P(j-q;j,3)-P(j-q+1;j,3)$ provides the following expression
\begin{equation}\label{eq:QjmqjN3}
 Q(j-q;j,3) = \frac{j-q}{3}+q_{3m}\text{ with }
 q_{3m}=\left(-\frac{1}{6},\frac{1}{2},\frac{1}{6}\right)
 \text{ for }j-q\bmod3=(1/2,3/2,5/2)\text{ respectively.}
\end{equation}
The expression (\ref{eq:PjmqjN3}) for $P(j-q;j,3)$ remains valid for $q=-1$, 
therefore the above expression applies if $q=0$. One verifies easily that 
(\ref{eq:QjpqjN3},\ref{eq:QjmqjN3}) are both correct for $q=0$.
One may also use the general expression (\ref{eq:PMjN3gen}). When computing the 
difference $P(j-q;j,3)-P(j-q+1;j,3)$ some attention must be paid to the case $q=0$ 
for which $H(q)=1\ne H(q-1)=0$. However the term in the factor of $H(q-1)$ is 
$(q-1)^2/4+\gamma(q-1)=0$ for $q=0$. The evaluation of $Q(j-q;j,3)$ is then 
straightforward, defining $\overline{\alpha}(n)=\alpha(n)-\alpha(n-1)$, 
$\overline{\gamma}(n)=\gamma(n)-\gamma(n-1)$. One gets 
\begin{subequations}\begin{align}\label{eq:QJjN3gen}
 Q(j-q;j,3)&=\frac16\left(2j+q-\frac12\right)+\overline{\alpha}(2j+q)
  -H(q)\left[\frac{q}{2}-\frac14+\overline{\gamma}(q)\right]\\
 \text{ with }\quad\overline{\alpha}(n)&=\left(\frac{1}{12},-\frac{1}{12},-\frac14,
 \frac{7}{12},-\frac{7}{12},\frac14\right)
 \text{for }n\bmod6=(0,1,2,3,4,5)\text{ respectively,}\\
 \overline{\gamma}(q)&=\left(\frac14,-\frac14\right)
 \text{ for }q\bmod2=(0,1) \text{ respectively,}
\end{align}\end{subequations}
with the conditions $3-2j\le q\le j-1/2$, since one must have $1/2\le j-q\le 3j-3$.

\section{Four-fermion systems}
\label{sec:PMj4}
\subsection{Determination of $P(M;j,4)$ if $M\ge2j$}
We first derive the expressions for $P(2j+p;j,4)$ which are easier to obtain than 
the expressions for $P(2j-p;j,4)$. One has for any natural integer $p$ 
\begin{subequations}\begin{align}
P(2j+p;j,4) &= P(2j+p;j-1,4)+P(j+p;j-1,3)+P(3j+p;j-1,3)+P(2j+p;j-1,2)\\
 &= P(2j+p;j-1,4)+P(j+p;j-1,3)\\
 &= P(2j+p;j-2,4)+P(j+p;j-1,3)+P(j+p+1;j-2,3)\\
 &= \sum_{s=1}^\sigma P(j+p+s-1;j-s,3)\label{eq:Sp2jpj4}
\end{align}\end{subequations}
where we used the properties $3j+p>J_\text{max}(j-1,3)$ and $2j+p>J_\text{max}(j-1,2)$. 
The upper bound $\sigma$ in (\ref{eq:Sp2jpj4}) is chosen so that $P(j+p+s-1;j-s,3)$ 
is nonzero if $s=\sigma$ and zero if $s=\sigma+1$, implying that $P(2j+p;j-\sigma,4)=0$. 
Explicitly
\begin{subequations}\begin{gather}
  j+p+\sigma-1\le J_\text{max}(j-\sigma,3)=3j-3\sigma-3\nonumber\\
  \text{and}\quad
  j+p+(\sigma+1)-2> J_\text{max}(j-(\sigma+1),3)=3j-3(\sigma+1)-3\\
  \sigma=\left\lfloor\frac{2j-p-2}{4}\right\rfloor\label{eq:sig}.
  \end{gather}
\end{subequations}
In order that the above formulas be meaningful one must have $\sigma\ge1$ or
\begin{equation}
 2j-p\ge6\quad\text{ from which }2j+p\le4j-6=J_\text{max}(j,4)\label{eq:condjp}.
\end{equation}
The sum (\ref{eq:Sp2jpj4}) will be calculated with formulas (\ref{eq:PjpqjN3}). 
This lead us to define $d(s)=j'-q'/2$ with $j'=j-s, q'=p+2s-1$, or
\begin{equation}\label{eq:defds} d(s)=j-\frac{p}{2}-2s+\frac12.\end{equation}
The following analysis will be done according to the value of 
$d(1)=j-p/2-3/2$. From (\ref{eq:sig}) 
\begin{equation}
\sigma=\left\lfloor\frac12d(1)+\frac14\right\rfloor
=\left\lfloor\frac12\left(j-\frac{p}{2}-\frac32\right)+\frac14\right\rfloor.
\end{equation}
To describe the procedure used to get $P(2j+p;j,4)$ let us consider the case 
$d(1)=j-p/2-3/2=3n$, where $n$ is an integer. One has then $\sigma=\lfloor3n 
+1/4\rfloor$ so that one must split the cases $n$ even and odd. If $n=2\nu$ 
with $\nu$ integer, then $\sigma=3\nu$. Writing $\theta(d)$ for the number on the 
right of the bracket in (\ref{eq:PjpqjN3}), we have
\begin{subequations}\begin{equation}
P(2j+p;j,4) = \sum_{s=1}^\sigma\frac13\left(j-\frac{p}{2}-2s+\frac12\right)^2 
+ \sum_{s=1}^\sigma \theta(d(s)).
\end{equation}
The quantity $d(s)=j-p/2-2s+1/2$ is equal to $j-p/2-3/2=3n=6\nu, 3n-2, 3n-4, 
\dots 2$ for $s=1\dots\sigma$.
Since $\sigma=3\nu$, there are $\nu$ elements in the sum such that $d\bmod3=0$, 
and as many such that $d\bmod3=1$ and  $d\bmod3=2$. The sum of $\theta(d)$ is 
according to (\ref{eq:PjpqjN3}), $(0-1/3-1/3)\nu=-2\nu/3$. The final result is
\begin{align}
P(2j+p;j,4) &= \sum_{t=1}^\sigma\frac43 t^2-\frac23\nu\\
 &= \frac{2}{9}(3\nu)(3\nu+1)(6\nu+1)-\frac23\nu=6\nu^2(2\nu+1)\\
 &=\frac{1}{18}\left(j-\frac{p}{2}-\frac32\right)^2\left(j-\frac{p}{2}+\frac32\right)
\text{ in the case }j-\frac{p}{2}-\frac32=6\nu.\label{eq:P2jpj4d6n0p}
\end{align}\end{subequations}
The procedure must be repeated in the cases $d(1)=3n,3n+1/2,3n+1,3n+3/2,3n+2,3n+5/2$, 
with $n=2\nu$, $n=2\nu+1$ where $\nu$ is integer.
From the expression (\ref{eq:PjpqjN3}), one notes that 
the sought number $P(M;j,4)$ is a sum of $P(j+t;j,3)$ that can be written as
\begin{subequations}\begin{align}\label{eq:P2jpj4s}
P(2j+p;j,4)&=\frac13\sum_{s=1}^\sigma [d(s)]^2-\sum_{s=1}^\sigma\theta(d(s))\\
&=\frac43\sum_{s=1}^\sigma \left(\frac{j}{2}-\frac{p-1}{4}-s\right)^2
 -\sum_{s=1}^\sigma\theta(d(s))\\
&=\frac43\sum_{t=0}^{\sigma-1}\left(t+b\right)^2
 -\sum_{s=1}^\sigma\theta(d(s))\label{eq:P2jpj4}
\end{align}\end{subequations}
where $b=j/2-(p-1)/4-\sigma$ is the smallest value of the 
quantity $d(s)/2$ in the sum (\ref{eq:P2jpj4s}).
Therefore the computation of $P(2j+p;j,4)$ amounts to 
obtaining the sum of the squares of numbers in arithmetical progression, which is 
easy to evaluate. This sum must be corrected by the term $\sum_{s}\theta(d(s))$. 

\begin{table}[htb]
\centering\renewcommand*{\arraystretch}{1.5}
\begin{tabular}{c@{\quad}c@{\quad}c@{\quad}c@{\quad}c@{\quad}c@{\quad}c}
\hline\hline
$j-\frac{p}{2}-\frac{3}{2}$ & $6\nu$ & $6\nu+1$ & $6\nu+2$ & $6\nu+3$ & $6\nu+4$ & $6\nu+5$\\
\hline
 $\sigma$ & $3\nu$ & $3\nu$ & $3\nu+1$ & $3\nu+1$ & $3\nu+2$ & $3\nu+2$ \\
 $b$    & $1$ & $3/2$ & $1$ & $3/2$ & $1$ & $3/2$\\ 
$\sum_s\theta(d(s))$ & $-2\nu/3$ & $-2\nu/3$ & $-(2\nu+1)/3$ & $-2\nu/3$ & $-2(\nu+1)/3$ & $-(2\nu+1)/3$ \\
$P(2j+p;j,4)$& $6\nu^2(2\nu+1)$ & $3\nu(2\nu+1)^2$ & $e_2$ & $3(\nu+1)(2\nu+1)^2$ & $6(\nu+1)^2(2\nu+1)$ & $e_5$ \\
\hline\hline
\end{tabular}
\caption{Various cases for the computation of $P(2j+p;j,4)$ in the case 
$p=2q$ even, i.e., $j-p/2+1/2$ integer. The number of states $P(2j+p;j,4)$ is 
given by $\frac43\sum_{t=0}^{\sigma-1}(t+b)^2+S$ with $S=\sum_s\theta(d(s))$.
The number of terms in the sum defining $P(2j+p;j,4)$ is 
$\sigma=\lfloor d(1)/2+1/4\rfloor$. One has $e_2=(2\nu+1)(6\nu^2+6\nu+1)$, 
$e_5=(\nu+1)(12\nu^2+24\nu+11)$ for $j-p/2-3/2=6\nu+2$, $6\nu+5$ respectively.
\label{tab:calcP2jq4e}
}
\end{table}

\begin{table}[htb]
\centering\renewcommand*{\arraystretch}{1.5}
\begin{tabular}{c@{\quad}c@{\quad}c@{\quad}c@{\quad}c@{\quad}c@{\quad}c}
\hline\hline
$j-\frac{p}{2}-\frac{3}{2}$ & $6\nu+1/2$ & $6\nu+3/2$ & $6\nu+5/2$ & 
 $6\nu+7/2$ & $6\nu+9/2$ & $6\nu+11/2$ \\
\hline
$\sigma$ & $3\nu$ & $3\nu+1$ & $3\nu+1$ & $3\nu+2$ & $3\nu+2$ & $3\nu+3$\\
$b$ & $5/4$ & $3/4$ & $5/4$ & $3/4$ & $5/4$ & $3/4$ \\
$\sum_s\theta(d(s))$ & $\nu/12$ & $(\nu+3)/12$ & $(\nu-1)/12$ & $(\nu+2)/12$
 & $(\nu+2)/12$ & $(\nu+1)/12$ \\
$P(2j+p;j,4)$ & $\nu(12\nu^2+9\nu+2)$ & $o_1$ & $o_2$ & 
 $(\nu+1)(12\nu^2+15\nu+5)$& $3(\nu+1)^2(4\nu+3)$ & $3(\nu+1)^2(4\nu+5)$\\
\hline\hline
\end{tabular}
\caption{Various cases for the computation of $P(2j+p;j,4)$ in the case $p=2q+1$ 
odd, i.e., $j-p/2+1/2$ half-integer. One has $o_1=12\nu^3+15\nu^2+6\nu+1$ for 
$j-p/2-3/2=6\nu+3/2$ and $o_2=12\nu^3+21\nu^2+12\nu+2$ for $j-p/2-3/2=6\nu+5/2$.
See Table \ref{tab:calcP2jq4e} for details.\label{tab:calcP2jq4o}
}
\end{table}

The parameters $\sigma,b,\sum_{s}\theta(d(s))$  corresponding to each case are 
described in the Tables \ref{tab:calcP2jq4e} and \ref{tab:calcP2jq4o} for 
$p$ even and odd respectively. The last line of these tables provides the 
number of states as given by (\ref{eq:P2jpj4}). The $\nu$-dependent values 
can be expressed back versus the physical quantities $j,p$. With the 
additional definition 
\begin{equation}X=j-\frac{p}{2}-\frac{1}{2}\label{eq:defX}\end{equation}
the expressions for $P(2j+p;j,4)$ will be even simpler. 
In the case $p=2q$ even, from table \ref{tab:calcP2jq4e} results, a detailed 
inspection proves that the expression of $P(2j+2q;j,4)$ versus $j,q$ is identical 
for each pair of adjacent columns. Namely, columns $6\nu$ (resp. $6\nu+1$, $6\nu+2$) 
and $6\nu+3$ (resp. $6\nu+4$, $6\nu+5$) provide the same result, so that, the 
$P(2j+2q;j,4)$ value does not depend on $j-q-1/2\bmod6$ but on $j-q-1/2\bmod3$.
Expressing $\nu$ versus $j,q$ one has, using the definition (\ref{eq:defX}),
\begin{equation}\label{eq:P2jpj4e}
P(2j+2q;j,4)=\begin{cases}
 \frac{1}{18}\left(j-q-\frac32\right)^2\left(j-q+\frac32\right)
 =\frac{X^3}{18}-\frac{X}{6}+\frac19 &\quad\text{if }j-q+\frac32=3n\\
 \frac{1}{18}\left(j-q-\frac52\right)\left(j-q+\frac12\right)^2
 =\frac{X^3}{18}-\frac{X}{6}-\frac19 &\quad\text{if }j-q+\frac32=3n+1\\
 \frac{1}{18}\left(j-q-\frac12\right)\left[\left(j-q-\frac12\right)^2-3\right]
 =\frac{X^3}{18}-\frac{X}{6} &\quad\text{if }j-q+\frac32=3n+2.
\end{cases}
\end{equation}
In the case of odd $p$ one must also express $P(2j+2q+1;j,4)$ versus $j,q$ or more 
precisely versus $X=j-q-1$. The six cases considered in Table \ref{tab:calcP2jq4o} 
provide as many different expressions. As seen on Eq.~(\ref{eq:P2jpj4e}), the final 
expressions are simpler as functions of $X$. One has
\begin{align}\label{eq:P2jpj4o}
P(2j+2q+1;j,4)&=
\frac{X^3}{18}-\frac{X}{24}+\psi\left(j-q-\frac{1}{2}\right)\nonumber\\
\text{ where }\psi\left(j-q-\frac{1}{2}\right)&=\left(-\frac{1}{72},\frac{1}{72},
-\frac{1}{8},\frac{17}{72},-\frac{17}{72},\frac{1}{8}\right) 
\quad\text{if }j-q-\frac{1}{2}\bmod6=(0,1,2,3,4,5)\text{ respectively.}
\end{align}

\subsection{Determination of $P(M;j,4)$ if $M<2j$}
From Talmi's equation one has, assuming $q$ positive integer
\begin{equation}
  P(2j-q;j,4)=P(2j-q;j-1,4)+P(3j-q;j-1,3)+P(j-q;j-1,3)+P(2j-q;j-1,2)
\end{equation}
which suggests implementing a recurrence on $q$. Indeed the elements $P(3j-q;j-1,3),
P(j-q;j-1,3),P(2j-q;j-1,4)$ are known. In addition 
\begin{equation}P(2j-q,j-1,4)=P(2(j-1)-(q-2),j-1;4)\end{equation}
shows that the expression for $P(2j+1,j;4)$ (resp. $P(2j,j;4)$) obtained 
above --- using (\ref{eq:P2jpj4o}), (\ref{eq:P2jpj4e}) respectively ---, allows us 
to get $P(2j-1;j,4)$ (resp. $P(2j-2;j,4)$). 
This leads us to split the discussion according to the parity of $q$.
We first define
\begin{equation} F(j,q)=P(3j-q;j-1,3)+P(2j-q;j-1,2).\label{eq:defF}\end{equation}
Using the expression (\ref{eq:PjpqjN3}), it is easy to prove that, if $j\ge q+1/2$, 
\begin{equation} F(j,2q)=\begin{cases}
  q^2/3 & \text{ if } q\bmod3=0\\ (q^2-1)/3 & \text{ if } q\bmod3=1, 2,
 \end{cases}\label{eq:Fj2q}\end{equation}
and, assuming again $j\ge q+1/2$, that
\begin{equation} F(j,2q+1)=\begin{cases}
  q(q+1)/3 & \text{ if } q\bmod3=0, 2\\ (q^2+q+1)/3 &\text{ if } q\bmod3=1.
 \end{cases}\label{eq:Fj2q1}\end{equation}

\subsubsection{Computation of $P(2j-2p;j,4)$}
We first consider the case where $q$ is even.
The above formula for $F(j,2p)$ provides us with the expression for 
$P(2j-2;j,4)$. Using (\ref{eq:PjmqjN3}) for $P_3=P(j-q,j-1,3)$ and 
(\ref{eq:P2jpj4e})
for $P_1=P(2j-2;j-1,4)$ we get $P(2j-2;j,4)=P_1+P_3$. Writing $x=j-3/2$, 
one considers three cases according to $j\bmod3$.
\begin{itemize}\begin{subequations}\label{eq:P2jm2j4}
\item[$\bullet$]If $j\bmod3=1/2$, $P_1=x^3/18-x/6-1/9, 
 P_3=2(j-2)+\frac13(j-7/2)^2$, so that
\begin{equation}P(2j-2;j,4)=\frac{1}{144}\left(8j^3+12j^2-16j+5\right)
=\frac{(j+1/2)^3}{18}-\frac{j+1/2}{6}+\frac19.
\end{equation}
\item[$\bullet$]If $j\bmod3=3/2$, $P_1=x^3/18-x/6, 
 P_3=2(j-2)+\frac13(j-9/2)(j-5/2)$, whence
\begin{equation}P(2j-2;j,4)=\frac{1}{144}\left(8j^3+12j^2-16j-27\right)
=\frac{(j+1/2)^3}{18}-\frac{j+1/2}{6}-\frac19.
\end{equation}
\item[$\bullet$]If $j\bmod3=5/2$, $P_1=x^3/18-x/6+1/9, 
 P_3=2(j-2)+\frac13(j-9/2)(j-5/2)$, from which
\begin{equation}P(2j-2;j,4)=\frac{1}{144}\left(8j^3+12j^2-16j-11\right)
=\frac{(j+1/2)^3}{18}-\frac{j+1/2}{6}.
\end{equation}\end{subequations}
\end{itemize}
A series of similar computations for greater values of $q$ has been performed 
and leads us to propose the formula
\begin{equation}
  P(2j-2q;j,4)=\frac{(j+q-1/2)^3}{18}-\frac{(j+q-1/2)}{6}
  -f(q)+\theta(j+q)\label{eq:hypP2jm2q}
\end{equation}
which we will prove by recurrence on $q$. The initial computations 
show that $f(0)=0, f(1)=0$, and the general expression for $f(q)$ will be 
obtained below. The initial value $q=0$ (\ref{eq:P2jm2j4}) requires that 
\begin{equation} \theta(j+q)=\begin{cases}
  -1/9 & \text{ if } j+q+1/2\bmod3=0\\
 0 & \text{ if } j+q+1/2\bmod3=1\\
 +1/9 & \text{ if } j+q+1/2\bmod3=2. \end{cases}\label{eq:theta}
\end{equation}

Let us assume the recurrence (\ref{eq:hypP2jm2q}) true up to $q=p$ (e.g., 
$p=0$ or $1$), and  prove it for $q=p+1$. With definition (\ref{eq:Fj2q})
\begin{equation}\label{eq:Fj2p2}
 F(j,2p+2)=\frac13(p+1)^2+\varepsilon_{p+1},\text{ with }
  \varepsilon_{p+1}=-\frac13\text{ if }p+1\bmod3=1 \text{ or } 2,
  \text{ otherwise }0,
\end{equation}
we get from the fundamental relation (\ref{eq:Talmi})
\begin{equation}\label{eq:Pj2p}
 P(2j-2p-2;j,4)=P(2(j-1)-2p;j-1,4)+P(j-2p-2;j-1,3)+F(j,2p+2).
\end{equation}
The recurrence hypothesis applies to the first term of (\ref{eq:Pj2p})
\begin{equation}
 P(2(j-1)-2p;j,4)=\frac{(j+p-3/2)^3}{18}-\frac{(j+p-3/2)}{6}-f(p)
  +\theta(j-1,p).
\end{equation}
The second term of (\ref{eq:Pj2p}) is obtained from (\ref{eq:PjmqjN3})
\begin{subequations}\begin{align}
 P(j-2p-2;j-1,3)&=\frac{1}{3}(j+p-1/2)^2-(p+1/2)^2-\frac{1}{12}+\tau(j+p)\\
\text{ with }
 \tau(j+p)&=\frac13 \text{ if }j-2p-\frac12\bmod3 = j+p-\frac12\bmod3 = 0, \\
\text{ and }\tau(j+p)&=0 \text{ if }j-2p-\frac12\bmod3 = j+p-\frac12\bmod3 = 1 \text{ or } 2.
\end{align}\end{subequations}
In order to verify the recurrence for $q=2p+2$, according to Eq. 
(\ref{eq:Pj2p}) one must verify for every $j$
\begin{equation}\label{eq:cond_recpair}
 \Delta-f(p+1)+\theta(j+p+1)=-f(p)+\theta(j-1+p)+\varepsilon_{p+1}+\tau(j+p)
\end{equation}
where $\Delta$ contains the terms function of $j,p$ except $f(p)$ and the 
quantities defined modulo 3
\begin{align}\nonumber
 \Delta=\frac{(j+p+1/2)^3}{18}-\frac{(j+p+1/2)}{6}&-\frac{(j+p-3/2)^3}{18}
 +\frac{(j+p-3/2)}{6}\\&-\frac{1}{3}(j+p-1/2)^2+(p+1/2)^2+\frac{1}{12}
 -\frac{(p+1)^2}{3}.
\end{align}
After some basic algebraic manipulations one obtains
\begin{equation} \Delta=\frac23p^2+\frac{p}{3}-\frac29.\end{equation}
In addition, one may verify that $\delta=\theta(j+p+1)-\theta(j-1+p)
-\tau(j+p)$ does not depend on $j$. Indeed
\begin{equation}
 \delta=\left\{\begin{matrix}1/9\\-1/9\\0\end{matrix}\right.
  -\left\{\begin{matrix}-1/9\\0\\1/9\end{matrix}\right.
  -\left\{\begin{matrix}1/3\\0\\0\end{matrix}\right.
  \text{\quad if }j+p-\frac12\bmod3=
   \left\{\begin{matrix}0\\1\\2\end{matrix}\right.
\end{equation}
which leads to $\delta=-1/9$ in all cases. Equation (\ref{eq:cond_recpair}) 
may be rewritten, using $\varepsilon_{p+1}$ as given by (\ref{eq:Fj2p2})
\begin{subequations}\begin{align}
 f(p+1)-f(p)&=\Delta+\delta-\varepsilon_{p+1}\\
  &=\frac23p^2+\frac{p}{3}-\frac13
  \begin{cases}+\frac13 &\text{ if }p\bmod3 = 0\text{ or }1\\
    &\text{ if }p\bmod3 = 2\\
  \end{cases}
\end{align}\end{subequations}
which is
\begin{equation}\label{eq:diffp}
f(p+1)-f(p)= 
\begin{cases}\frac13p(2p+1) &\text{ if }p\bmod3 = 0\text{ or }1\\
  \frac13(p+1)(2p-1)=\frac13p(2p+1)-\frac13 &\text{ if }p\bmod3 = 2.
\end{cases}
\end{equation}
Since $f(p+1)-f(p)$ is indeed independent of $j$ the recurrence 
assumption (\ref{eq:hypP2jm2q}) is verified. The proof is completed by 
the determination of $f(p)$. Applying Eq.~(\ref{eq:diffp}) for 
$p,p+1,p+2$, we get, whatever $p\bmod3$,
\begin{equation}\label{eq:diff3p}
f(p+3)-f(p)=\frac{p}{3}(2p+1)+\frac{(p+1)}{3}(2p+3)
 +\frac{(p+2)}{3}(2p+5)-\frac13=2p^2+5p+4.
\end{equation}
From the known initial values $f(0)=f(1)=0$ using (\ref{eq:diffp}) 
one gets $f(2)=1$, and more generally $f(p)$
\begin{equation}
f(3n+p_0)-f(p_0)=\sum_{t=0}^{n-1}\left[2(3t+p_0)^2+5(3t+p_0)+4\right]
\end{equation}
and considering $p_0=0,1,2$ separately we obtain 
\begin{subequations}\begin{align}\label{eq:valfp}
f(p)&=f_0(p)= 
\frac{m}{2}(12m^2-3m-1)=\frac29p^3-\frac{p^2}{6}-\frac{p}{6} &\text{ if }p=3m\\
f(p)&=f_1(p)= \frac{m}{2}(12m^2+9m+1)=\frac29p^3-\frac{p^2}{6}-\frac{p}{6}+\frac19 
 &\text{ if }p=3m+1\\
f(p)&=f_2(p)= \frac{(m+1)}{2}(12m^2+9m+2)=\frac29p^3-\frac{p^2}{6}-\frac{p}{6}+\frac29 
&\text{ if }p=3m+2.
\end{align}\end{subequations}
A further generalization consists in verifying that the expression
(\ref{eq:hypP2jm2q}) may be applied \emph{even for $q$ negative} 
provided one cancels the $f(q)$ term. Indeed comparing this expression 
to the known values (\ref{eq:P2jpj4e}), one notes that 
\begin{equation}
 P(2j-2q;j,4)=\frac{(j+q-1/2)^3}{18}-\frac{(j+q-1/2)}{6}+\theta(j+p)
 \quad\text{if }q<0
\end{equation}
from which one gets, whatever the sign of the integer $q$,
\begin{equation}
  P(2j-2q;j,4)=\frac{(j+q-1/2)^3}{18}-\frac{(j+q-1/2)}{6}-H(q)f(q)
   +\theta(j+q)
\end{equation}
$H(q)$ being the Heaviside function, $H(q)=1$ if $q\ge0$, 0 otherwise.

\subsubsection{Computation of $P(2j-2p-1;j,4)$}
As a first example, the computation of $P(2j-1;j,4)$ is detailed in 
Appendix \ref{sec:P2jm1j4}. In order to discover the general formula, 
we also got expressions for $P(2j-3;j,4)$ and $P(2j-5;j,4)$.
An analysis on $P(M;j,4)$ with $2j-M$ odd similar to the case $2j-M$ 
even leads us to propose the relation
\begin{equation}\label{eq:hypP2jm2q1}
  P(2j-2q-1;j,4)=\frac{(j+q)^3}{18}-\frac{(j+q)}{24}
  -g(q)+\phi(j+q)
\end{equation}
which will be demonstrated by recurrence. The direct computation in the 
first two cases show that $g(0)=0$ and $g(1)=0$. From the analysis of 
Appendix \ref{sec:P2jm1j4} one imposes that, if $q\ge0$,
\begin{equation} \phi(j+q)=\left(\frac{1}{72},-\frac{1}{8},
\frac{17}{72},-\frac{17}{72},\frac{1}{8},-\frac{1}{72}\right)\text{ if }
j+q-\frac12\bmod6=(0,1,2,3,4,5)\text{ respectively.}\label{eq:valphi}
\end{equation}
Assuming that (\ref{eq:hypP2jm2q1}) is true up to $q=p$, we now try to 
prove it for $q=p+1$. Using the value (\ref{eq:Fj2q1})
\begin{equation}\label{eq:Fj2p3}
 F(j,2p+3)=\frac13(p+1)(p+2)+\mu_{p+1},\text{ where }
  \mu_{p+1}=\frac13\text{ if }(p+1)\bmod3=1, \text{ otherwise }\mu_{p+1}=0,
\end{equation}
the fundamental relation (\ref{eq:Talmi}) may be written
\begin{equation}\label{eq:Pj2p1}
 P(2j-2p-3;j,4)=P(2(j-1)-2p-1;j-1,4)+P(j-2p-3;j-1,3)+F(j,2p+3).
\end{equation}
The recurrence hypothesis is again applied to the first term at the second 
member of (\ref{eq:Pj2p1})
\begin{equation}
 P(2(j-1)-2p-1;j,4)=\frac{(j+p-1)^3}{18}-\frac{(j+p-1)}{24}-g(p)
  +\phi(j-1,p).
\end{equation}
The second term at the second member of (\ref{eq:Pj2p1}) is given by 
(\ref{eq:PjmqjN3s})
\begin{subequations}\begin{align}
 P(j-2p-3;j-1,3)&=\frac{1}{3}(j-1+p+1)^2-(p+1)^2-\frac{1}{12}+\upsilon(j+p)\\
\text{ with }
 \upsilon(j+p)&=\frac13 \text{ if  }j-2p-\frac12\bmod3 = j+p-\frac12\bmod3 = 1, \\
\text{ and }\upsilon(j+p)&=0 \text{ if }j-2p-\frac12\bmod3 = j+p-\frac12\bmod3 = 0 \text{ or } 2.
\end{align}\end{subequations}
In order to verify the recurrence for $q=2p+3$, from (\ref{eq:Pj2p1}) 
one must have for every $j$
\begin{equation}\label{eq:cond_recimpair}
 \Delta'-g(p+1)+\phi(j+p+1)=-g(p)+\phi(j-1+p)+\mu_{p+1}+\upsilon(j+p)
\end{equation}
where $\Delta'$ contains the terms function of $j,p$ except $g(p)$ and the 
terms defined modulo 3 or modulo 6
\begin{equation}
 \Delta'=\frac{(j+p+1)^3}{18}-\frac{(j+p+1)}{24}-\frac{(j+p-1)^3}{18}
 +\frac{(j+p-1)}{24}-\frac{1}{3}(j+p)^2+(p+1)^2+\frac{1}{12}-\frac13(p+1)(p+2).
\end{equation}
After some algebra one gets
\begin{equation} \Delta'=\frac23p^2+p+\frac49.\end{equation}
In addition one can check that $\delta'=\phi(j+p+1)-\phi(j-1+p)
-\upsilon(j+p)$ is independent of $j$. Indeed
\begin{equation} \delta'=
\left\{\begin{matrix}-1/8\\17/72\\-17/72\\1/8\\-1/72\\1/72\end{matrix}\right.
-\left\{\begin{matrix}-1/72\\1/72\\-1/8\\17/72\\-17/72\\1/8\end{matrix}\right.
-\left\{\begin{matrix}0\\1/3\\0\\0\\1/3\\0\end{matrix}\right.
  \text{\quad if }j+p-\frac12\bmod3=
   \left\{\begin{matrix}0\\1\\2\\3\\4\\5\end{matrix}\right.
\end{equation}
therefore $\delta'=-1/9$ in all cases. Equation (\ref{eq:cond_recimpair}) 
may be rewritten as 
\begin{equation}
 g(p+1)-g(p)=\Delta'+\delta'-\mu_{p+1}=\frac23p^2+p+\frac13-\mu_{p+1}
\end{equation}
which is, using $\mu_{p+1}$ as given by (\ref{eq:Fj2p3}),
\begin{equation}\label{eq:difgp}
g(p+1)-g(p)= 
\begin{cases}
\frac13p(2p+3)=\frac13(p+1)(2p+1)-\frac13 &\text{ if }p\bmod3 = 0\\
\frac13(p+1)(2p+1) &\text{ if }p\bmod3 = 1\text{ or }2.
\end{cases}
\end{equation}
Since $g(p+1)-g(p)$ is indeed independent of $j$ the recurrence 
relation (\ref{eq:hypP2jm2q1}) is proved and $g(p)$ may be computed. 
One may use (\ref{eq:difgp}) in order to get $g(1)=0, g(2)=2$ from $g(0)=0$.
One has then, whatever $p\bmod3$,
\begin{equation}\begin{split}\label{eq:difg3p}
g(p+3)-g(p)=\frac{(p+1)}{3}(2p+1)+\frac{(p+2)}{3}(2p+3)
 +\frac{(p+3)}{3}(2p+5)-\frac13\\=2p^2+7p+7.
\end{split}\end{equation}
From the values of $g(p)$ for $p_0=0,1,2$ one obtains the general expression
\begin{equation}
g(3n+p_0)-g(p_0)=\sum_{t=0}^{n-1}\left[2(3t+p_0)^2+7(3t+p_0)+7\right]
\end{equation}
and splitting cases $p_0=0,1,2$,
\begin{subequations}\label{eq:valgp}\begin{align}
g(p)&=g_0(p)= 
\frac{n}{2}(12n^2+3n-1)=\frac29p^3+\frac{p^2}{6}-\frac{p}{6} &\text{ if }p=3n\\
g(p)&=g_1(p)= \frac{n}{2}(12n^2+15n+5)=\frac29p^3+\frac{p^2}{6}-\frac{p}{6}-\frac29 
 &\text{ if }p=3n+1\\
g(p)&=g_2(p)= \frac{(n+1)}{2}(12n^2+15n+4)=\frac29p^3+\frac{p^2}{6}-\frac{p}{6}-\frac{1}{9} 
&\text{ if }p=3n+2.
\end{align}\end{subequations}
As for even $p$, comparing expressions (\ref{eq:P2jpj4o}) and (\ref{eq:hypP2jm2q1}) 
for $q$ negative or positive, one may write the general relation
\begin{equation}
 P(2j-2q-1;j,4)= \frac{(j+q)^3}{18}-\frac{(j+q)}{24}-H(q)g(q)+\phi(j+q)
\end{equation}
where $H(q)$ is the Heaviside function, and $\phi(j+q)$ is given by 
(\ref{eq:valphi}).

\subsubsection{General expression for the distribution of the 
magnetic quantum number}
The formulas (\ref{eq:hypP2jm2q}), (\ref{eq:hypP2jm2q1}) for $P(2j-n;j,4)$ in the 
cases $n$ even and odd can even be gathered in a single expression. One 
notices that the first term can be simply written as $(j+(n-1)/2)^3/18$, 
while the second is $-(j+(n-1)/2)/6+\pi(n)(j+(n-1)/2)/8$ where $\pi(n)$ is 0 
if $n$ is even, 1 if $n$ is odd. The third term of the quoted formulas may 
also be unified, noting that from the values (\ref{eq:valfp},\ref{eq:valgp}), 
one has
\begin{equation}g_2(x-1/2)=f_1(x)-\frac18\end{equation}
which allows us to write $f(n/2)$ and $g((n-1)/2)$ with a single formula, 
namely $f(n/2)=f_1(n/2)+\xi(n)$ with $\xi(n)=(-1/9,0,1/9)$ if $n/2\bmod3=
0,1,2$ respectively, and $g((n-1)/2)=f_1(n/2)+\xi(n)$ with 
$\xi(n)=(-1/8+1/9,-1/8-1/9,-1/8)$ for $(n-1)/2\bmod3=0,1,2$ respectively. 
Finally the term $\theta(j+n/2)$ and $\phi(j+(n-1)/2)$ in these formulas can 
be collected in a single expression, if one considers $2j+n\bmod12$ value. If 
$n$ is even, from the expression (\ref{eq:theta}) one may write this term as 
$\theta(j+n/2)=\left(0,1/9,-1/9,0,1/9,-1/9\right)$ for $2j+n-1\bmod12=
(0,2,4,6,8,10)$ respectively. If $n$ is odd, from (\ref{eq:valphi}) this term 
is $\phi(j+q)=\phi(j+(n-1)/2)=(1/72,-1/8,17/72,-17/72, 1/8,-1/72)$ for 
$2j+n-1\bmod12=(1,3,5,7,9,11)$ respectively. One obtains the single formula
\begin{subequations}\label{eq:PMjN4gen}\begin{align}
P(2j-n;j,4)&=\frac{1}{18}\left(j+\frac{n-1}2\right)^3
 -\left(\frac16-\frac{\pi(n)}{8}\right)\left(j+\frac{n-1}2\right)
 -H(n)\left[f_1\left(\frac{n}2\right)+\xi(n)\right]+\omega(2j+n-1)\\
\text{with}\quad \pi(n)&=n\bmod2,\quad \label{eq:xi}
 \xi(n)=\left(-\frac{1}{9},-\frac{1}{72},0,-\frac{17}{72},\frac{1}{9},
 -\frac{1}{8}\right)\text{ if } n\bmod6=(0,1,2,3,4,5)\text{ respectively},\\
 f_1(n)&=\frac29n^3-\frac{n^2}{6}-\frac{n}{6}+\frac19,\label{eq:f1}\\
  \omega(2j+n-1)&=\nonumber
\left(0,\frac{1}{72},\frac{1}{9},-\frac{1}{8},-\frac{1}{9},\frac{17}{72},0,
-\frac{17}{72},\frac{1}{9},\frac{1}{8},-\frac{1}{9},-\frac{1}{72}\right)\\
&\quad\text{ if }2j+n-1\bmod12=(0,1,2,3,4,5,6,7,8,9,10,11)\text{ respectively}.
\label{eq:omega}\end{align}\end{subequations}
In this formula $2j-n$ must be non-negative, $n$ may be negative. Explicitly, 
$n$ must be such that $-2j+6\le n\le2j$.

\begin{figure}[htbp]
 \begin{minipage}[c]{0.49\textwidth}
   \centering
   \includegraphics[width=\textwidth,angle=\anglefig,scale=\scalefig]{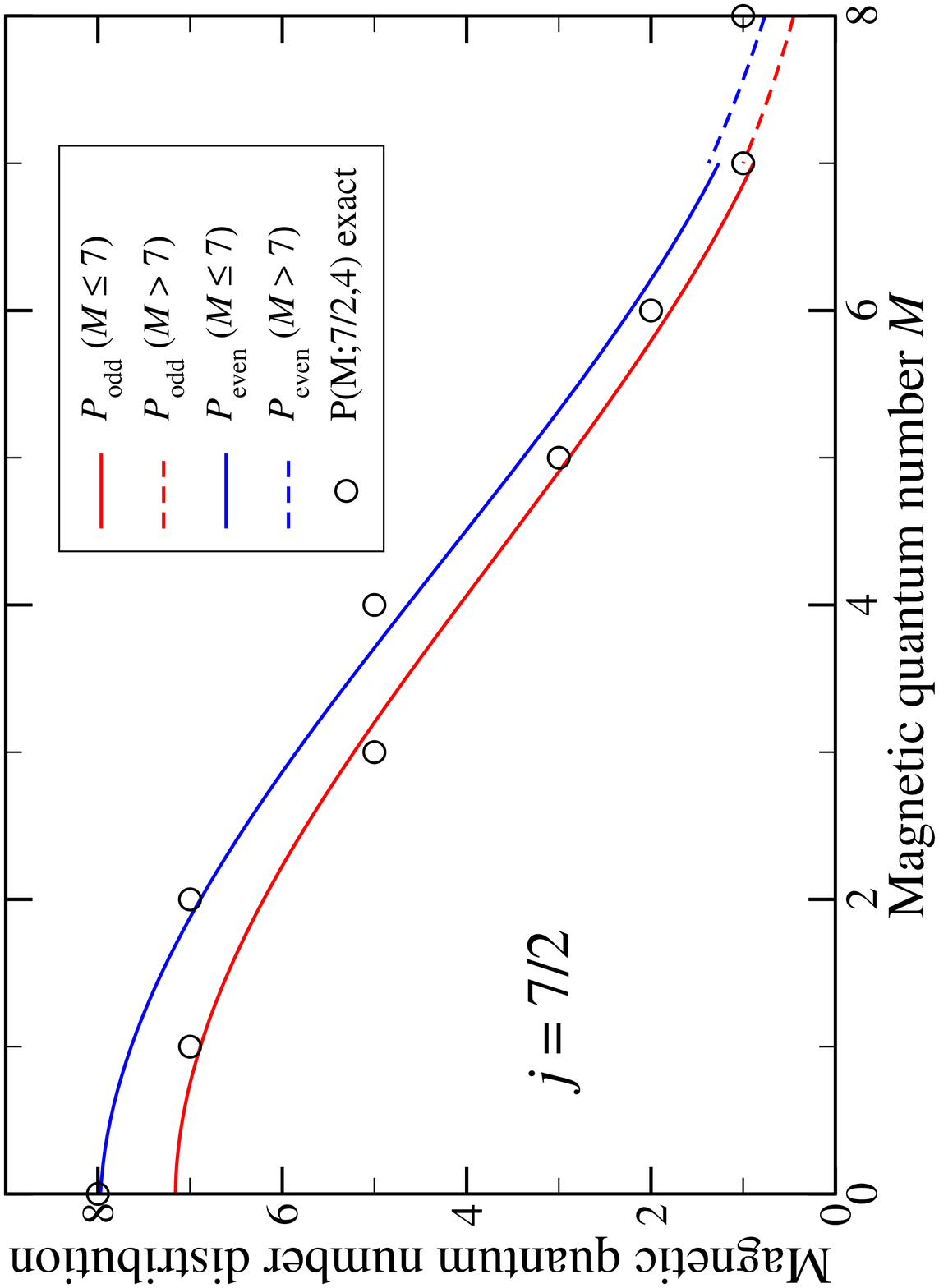}
   \caption{(Color online) Magnetic quantum number distribution $P(M;j,4)$ for a 
   four-fermion system with spin $j=7/2$. The red (resp. blue) curve is the 
   approximation (\ref{eq:PMjN4approx}) with $\pi(n)=0$ (resp. 1). The black circles
   are the exact values.\label{fig:PMj7s2N4}}
 \end{minipage}\hfill
 \begin{minipage}[c]{0.49\textwidth}
   \centering
   \includegraphics[width=\textwidth,angle=\anglefig,scale=\scalefig]{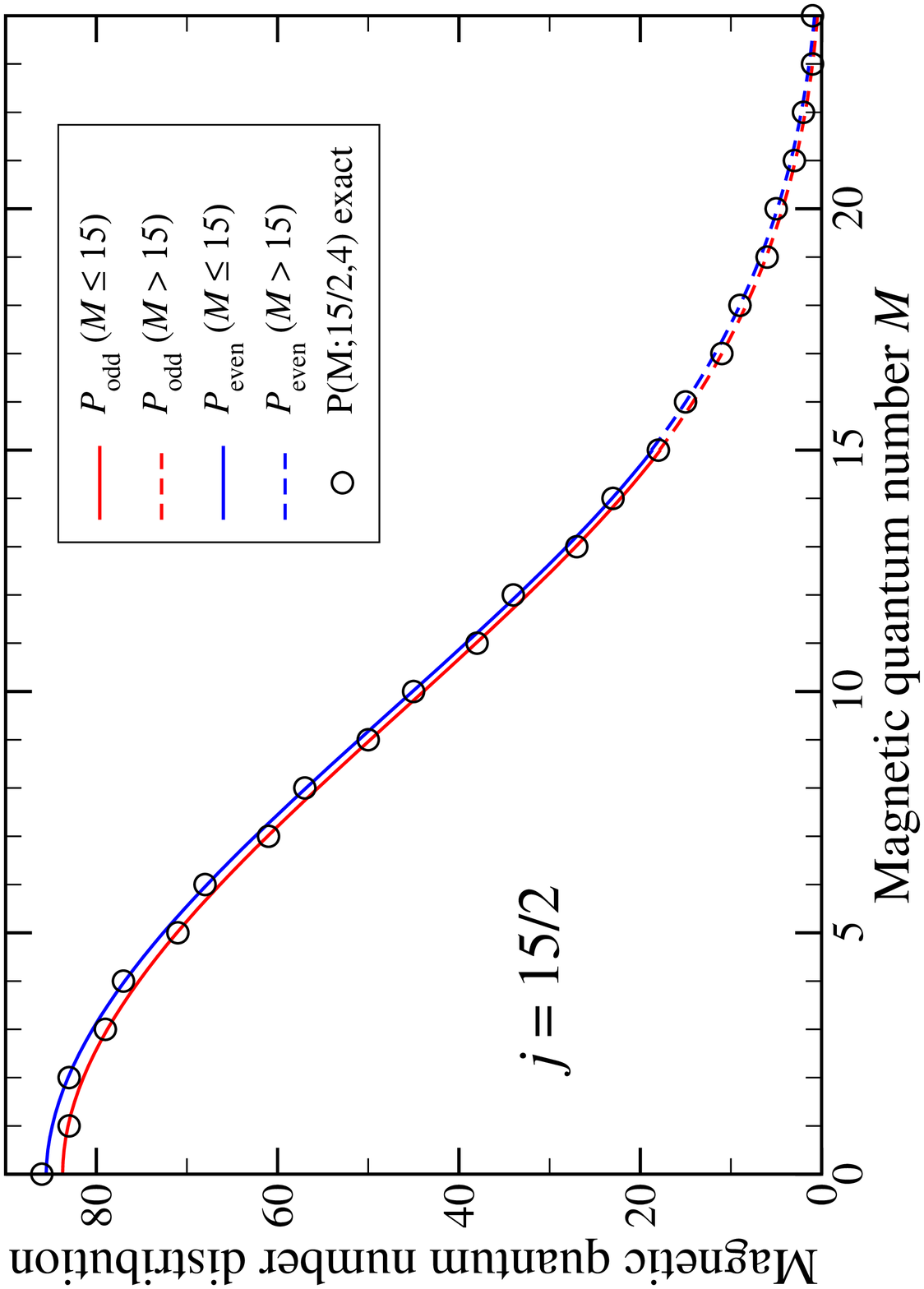}
   \caption{(Color online) Magnetic quantum number distribution $P(M;j,4)$ for a 
   four-fermion system with spin $j=15/2$. See Fig.~\ref{fig:PMj7s2N4} for details.
   \label{fig:PMj15s2N4}}\quad
 \end{minipage}
\end{figure}

Though this paper is not devoted to deriving approximations, one will observe that 
for $n\ge0$, one has $|\omega(2j+n-1)-\xi(n)|\le17/36$, and for $n<0$ one has 
$|\omega(2j+n-1)|\le17/72$, so that the approximation
\begin{equation}\label{eq:PMjN4approx}
P(2j-n;j,4)\simeq P_\text{app}(2j-n;j,4)=\frac{1}{18}\left(j+\frac{n-1}2\right)^3
 -\left(\frac16-\frac{\pi(n)}{8}\right)\left(j+\frac{n-1}2\right)
 -H(n)f_1\left(\frac{n}2\right)
\end{equation}
results in an absolute error below 1/2. The relative error will be small if 
conditions $(j+(n-1)/2)\gg1, n\gg 1$ are met. For really large $j$, even the 
$\pi(n)$ dependent term may be omitted, but the resulting approximation is not as 
good. This is illustrated by Figs.~\ref{fig:PMj7s2N4} and \ref{fig:PMj15s2N4} for 
$j=7/2$ and $15/2$ respectively. As can be seen in 
the approximate form above, both 
approximations $\pi(n)=0$ and 1 exhibit a discontinuity of 1/9 at $n=0$ or $M=2j$ 
since $f_1(0)=1/9$. Though the above approximation is rough for $j=7/2$ it proves to 
be fair for higher $j$, and correctly reproduces the even-odd staggering, previously 
noticed in the atomic physics context \cite{Bauche1997,Poirier2021b}.

\subsection{Total number of levels}
\label{subsec:Qtotj4}
A direct application of the above derived expression for $P(M;j,4)$ is the 
determination of the total number of levels. From the relation (\ref{eq:QvsP}), 
one verifies that the total number of levels for four fermions of spin $j$ is 
given by $P(0;j,4)$, which is easily obtained with (\ref{eq:hypP2jm2q1}). 
Writing $q=j-1/2$ in this equation, one gets
\begin{equation}\label{qtotj4}
 P(0;j,4)= \frac{(2j-1/2)^3}{18}-\frac{(2j-1/2)}{24}-g(j-1/2)+\phi(2j-1/2).
\end{equation}
One has to consider three cases according to $j-1/2\bmod3$. 
If $j-1/2=3n$, the first equation in the group (\ref{eq:valgp}) applies, and 
one has $\phi(2j-1/2)=1/72$. If $j-1/2=3n+1$, the second equation in the group 
(\ref{eq:valgp}) applies, and $\phi(2j-1/2)=17/72$. If $j-1/2=3n+1$, the third  
equation (\ref{eq:valgp}) is relevant, and $\phi(2j-1/2)=1/8$.
One obtains the general formula
\begin{equation}
 P(0;j,4)= \frac{2}{9}j^3-\frac{j^2}{6}+\frac{j}{6}
 \begin{cases}
 -5/72&\text{ if }j-1/2\bmod3=0,\\
 +3/8&\text{ if }j-1/2\bmod3=1,\\
 +11/72&\text{ if }j-1/2\bmod3=2.
 \end{cases}
\end{equation}

\subsection{Distribution of the total angular momentum}
\label{subsec:Qj4}
\begin{figure}[htbp]
 \begin{minipage}[c]{0.49\textwidth}
   \centering
   \includegraphics[width=\textwidth,angle=\anglefig,scale=\scalefig]{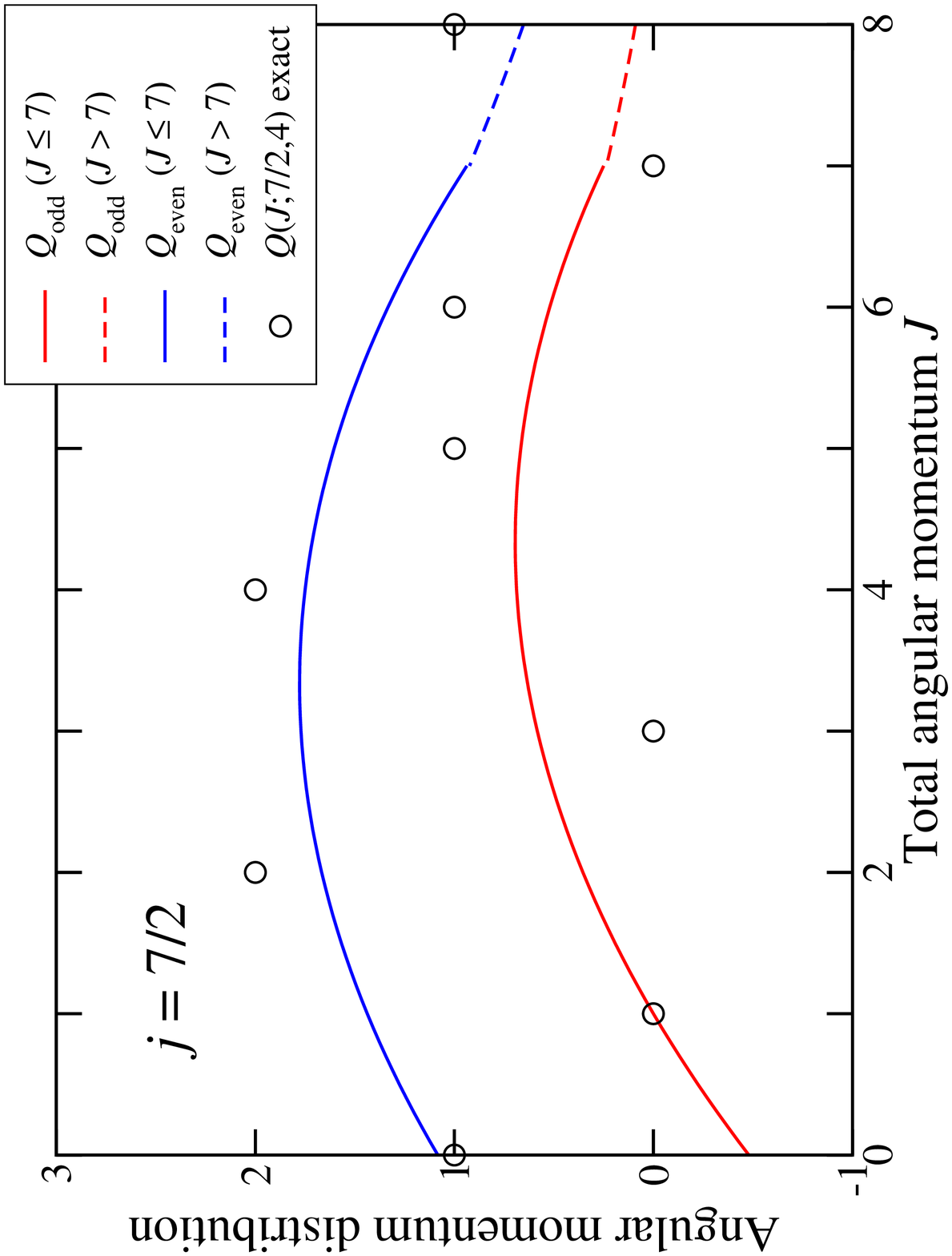}
   \caption{(Color online) Angular momentum distribution $Q(J;j,4)$ for a 
   four-fermion system with spin $j=7/2$. The red (resp. blue) curve is the 
   approximation (\ref{eq:QJjN4approx}) with $\pi(n)=0$ (resp. 1). The black circles
   are the exact values.\label{fig:QJj7s2N4}}
 \end{minipage}\hfill
 \begin{minipage}[c]{0.49\textwidth}
   \centering
   \includegraphics[width=\textwidth,angle=\anglefig,scale=\scalefig]{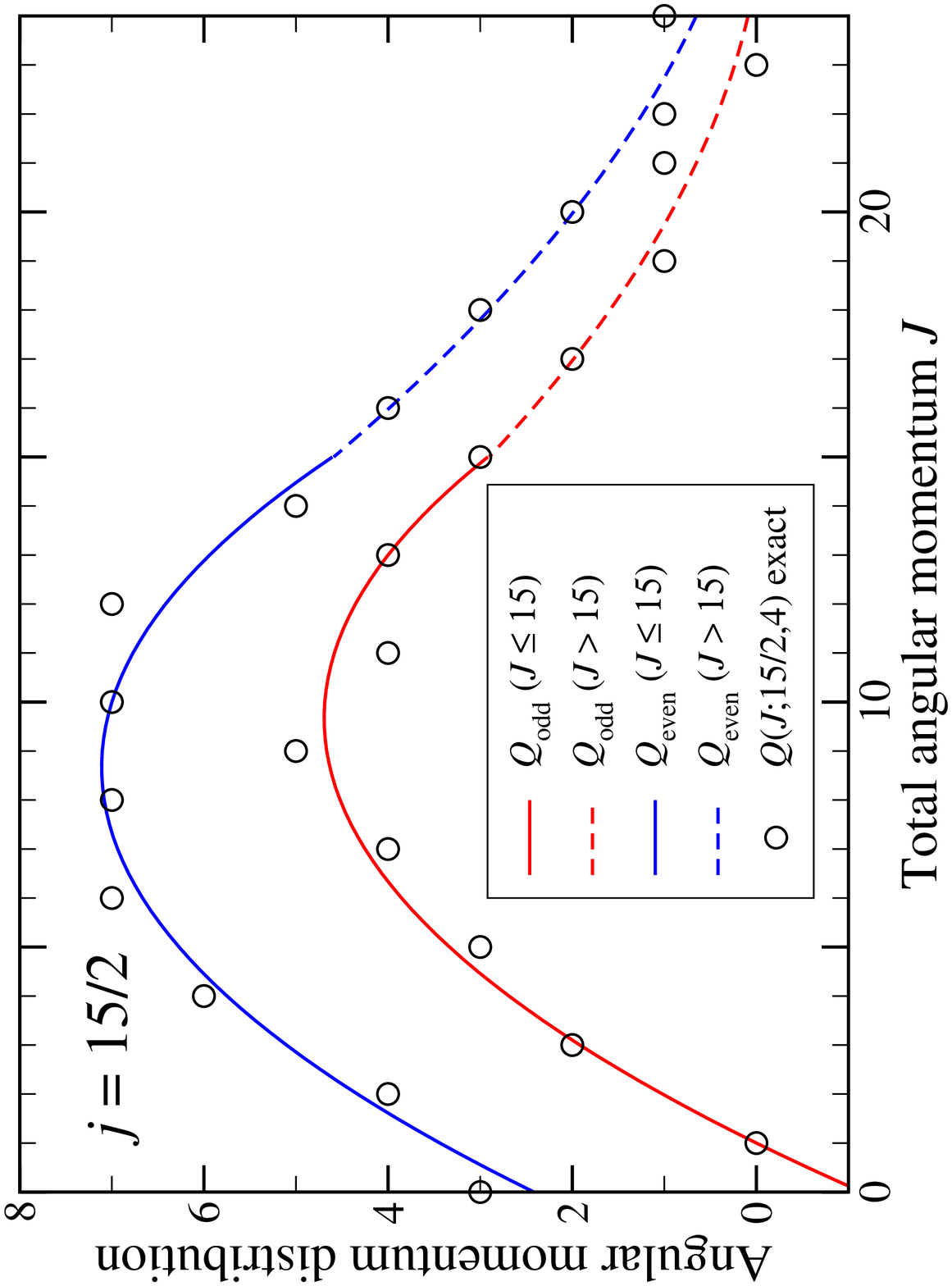}
   \caption{(Color online) Angular momentum distribution $Q(J;j,4)$ for a 
   four-fermion system with spin $j=15/2$. See Fig.~\ref{fig:QJj7s2N4} for details.
   \label{fig:QJj15s2N4}}\quad
 \end{minipage}
\end{figure}

Once again, the fundamental relation (\ref{eq:QvsP}), together with the expression 
(\ref{eq:PMjN4gen}) of the $M$ distribution for a four-fermion system, allow us to 
derive the distribution of the total momentum $J$. One must evaluate $Q(2j-n;j,4)=
P(2j-n;j,4)-P(2j-(n-1);j,4)$ which we will write as $Q_1+Q_2+Q_3$. The quantity 
$Q_1$ consists in the contribution of first two terms of (\ref{eq:PMjN4gen}), which 
is easily obtained noticing that $\pi(n-1)=1-\pi(n)$,
\begin{align}
Q_1&=\frac{X^3}{18}-\frac{X}{6}-\frac{(X-1/2)^3}{18}+\frac{X-1/2}{6}
 -\frac{j+n/2-1}{8}+\frac{\pi(n)}{8}(2j+n-3/2)\quad\text{ with }X=j+\frac{n-1}{2}\\
&=\frac{1}{12}\left(j+\frac{n-3}2\right)^2-\frac{7}{72}
 +\frac{\pi(n)}{8}(2j+n-3/2).
\end{align}
The quantity $Q_2$ is the difference of terms involving the Heaviside factors $H(n)$ 
and $H(n-1)$. These factors are equal except in the case $n=0$ which requires more 
attention: one must note that the factor of $H(n-1)$ for $n=0$ is $f_1(-1/2)+
\xi(-1)$, which is zero according to the values (\ref{eq:xi},\ref{eq:f1}). 
Therefore one may write 
\begin{subequations}\begin{align}
Q_2&=-H(n)\left[f_1(n/2)+\xi(n)-f_1((n-1)/2)-\xi(n-1)\right]\\
&=-H(n)\left[\frac{n^2}{12}-\frac{n}{6}-\frac{1}{72}+\overline{\xi}(n)\right]
\end{align}
with
\begin{equation}\overline{\xi}(n)=\xi(n)-\xi(n-1)=\left(\frac{1}{72},\frac{7}{72},
\frac{1}{72},-\frac{17}{72},\frac{25}{72},-\frac{17}{72}\right)
\text{ for }n\bmod6=0,1,2,3,4,5 \text{ respectively.}
\end{equation}\end{subequations}
Finally the $\omega$-dependent term is simply
\begin{subequations}\begin{align}
Q_3&=\overline{\omega}(2j+n-1)=\omega(2j+n-1)-\omega(2j+n-2)\\
&=\left(\frac{1}{72},\frac{1}{72},\frac{7}{72},-\frac{17}{72},\frac{1}{72},\frac{25}{72},
 -\frac{17}{72},-\frac{17}{72},\frac{25}{72},\frac{1}{72},-\frac{17}{72},\frac{7}{72}\right)
\end{align}\end{subequations}
for $2j+n-1\bmod12=0-11$ respectively. The complete formula is
\begin{align}\label{eq:Q2jmnjN4}
Q(2j-n;j,4)=&\frac{1}{12}\left(j+\frac{n-3}2\right)^2-\frac{7}{72}
 +\frac{\pi(n)}{8}(2j+n-3/2)\nonumber\\
 &-H(n)\left[\frac{(n-1)^2}{12}-\frac{7}{72}+\overline{\xi}(n)\right]
 +\overline{\omega}(2j+n-1).
\end{align}
Similarly to the $M$-distribution study, one observes that for $n\ge0$ one has 
$|\overline{\omega}(2j+n-1)-\overline{\xi}(n)|\le7/12$, while for $n<0$ one has 
$|\overline{\omega}(2j+n-1)|\le25/72$, so that the congruence-free approximation
\begin{equation}\label{eq:QJjN4approx}
Q(2j-n;j,4)\simeq Q_\text{app}(2j-n;j,4)=
\frac{1}{12}\left(j+\frac{n-3}2\right)^2-\frac{7}{72}+\frac{\pi(n)}{8}(2j+n-3/2)
-H(n)\left[\frac{(n-1)^2}{12}-\frac{7}{72}\right]
\end{equation}
holds with an error less than unity. The approximation is tested in Figs. 
\ref{fig:QJj7s2N4} and \ref{fig:QJj15s2N4} for $j=7/2$ and 15/2 respectively. 
Since the main contribution to $Q(J;j,4)$ scales as the squares $j^2$ or $n^2$ 
instead of cubes in the $P(M;j,4)$ case, the above approximation is not as good 
as $P_\text{app}(M;j,4)$. Nevertheless the above formula is quite simple and 
efficient for moderate $j$ values. As for the above $P(M;j,4)$ analysis, one notices 
a significant even-odd staggering \cite{Bauche1997,Poirier2021b}  which 
is correctly reproduced by the above formula. Finally one will note that 
the discontinuity on the approximate values at $n=0$ is only 1/72 so that the red 
and blue curves look almost continuous at $J=2j$.

\section{Total number of levels in five-fermion systems}
\label{sec:Qtotj5}
The formula (\ref{eq:PMjN4gen}) allows us to get the total number of levels 
for a five-fermion system, which is equal to $P(1/2;j,5)$. From (\ref{eq:Talmi}),
one may write, for $s$ from 1 to $s=j-3/2$,
\begin{equation}
P(1/2;j-s+1,5)=P(1/2;j-s,5)+P(j-s+3/2;j-s,4)+P(j-s+1/2;j-s,4)+P(1/2;j-s,3).
\end{equation}
which gives the total number of levels as a sum
\begin{subequations}\label{eq:S1S2S3}\begin{align}
 P(1/2;j,5)&=S_1+S_2+S_3\\
 S_1&=\sum_{s=1}^{j-5/2}P(j-s+3/2;j-s,4)\\
 S_2&=\sum_{s=1}^{j-5/2}P(j-s+1/2;j-s,4)\\
 S_3&=\sum_{s=1}^{j-3/2}P(1/2;j-s,3)
\end{align}\end{subequations}
knowing that for $s=j-3/2$ the elements $P(j-s+1\pm1/2;j-s,4)$ vanish. The sum 
$S_3$ is easily derived from (\ref{qtotj3})
\begin{equation}
S_3=\sum_{s=1}^{j-3/2}\frac12\left[(j-s)^2-\frac14\right]
 =\frac{1}{48}(2j-3)(2j-1)(2j+1)
 =\frac{j^3}{6}-\frac{j^2}{4}-\frac{j}{24}+\frac{1}{16}.
\end{equation}
Using the formula for the four-fermion distribution (\ref{eq:PMjN4gen}), the sum 
(\ref{eq:S1S2S3}) may be rewritten by gathering the contributions to $S_1$ and $S_2$
\begin{subequations}\begin{align}
 P(1/2;j,5)&=A-\Xi+\Omega\\
 A&=\sum_{u=0,1}\sum_{s=1}^{j-5/2}\left[\frac{(3j-3s-u-1/2)^3}{144}
  -\frac{(3j-3s-u-1/2)}{12}-f_1\left(\frac{j-s-u-1/2}{2}\right)\right]
  \nonumber\\
  &\quad+\sum_{u=0,1}\sum_{s=1}^{j-5/2}\left[\pi(j-s-u-1/2)\frac{(3j-3s-u-1/2)}{16}\right]+S_3\\
 \Xi&=\sum_{u=0,1}\sum_{s=1}^{j-5/2}\xi(j-s-u-1/2)\label{eq:defXi}\\
 \Omega&=\sum_{u=0,1}\sum_{s=1}^{j-5/2}\omega(3j-3s-u-3/2).\label{eq:defOmega}
\end{align}\end{subequations}
In order that $P(j-s+1\pm1/2;j-s,4)$ be nonzero, on must have $j\ge7/2$. When 
evaluating the last part of the sum $A$, because of the factor $\pi(j-s-u-1/2)$ one 
must consider separately the cases $j-1/2$ even and odd. We define $\overline{n}=j-s-u-1/2$. 
If $j-1/2=2\nu$ is even, we have $\overline{n}=2\nu-1-s$ (resp. $2\nu-s$) for $u=1$ 
(resp. $u=0$), and $\overline{n}$ will be odd if $s=2t, 1\le t\le(j-5/2)/2=\nu-1$ 
(resp. $s=2t-1, 1\le t\le (j-5/2)/2=\nu-1$). If $j-1/2=2\nu+1$ is odd, we have 
$\overline{n}=2\nu-s$ (resp. $2\nu+1-s$) for $u=1,0$ respectively, and since 
$\overline{n}$ must be odd, the summation index is $s=2t-1, 1\le t\le (j-3/2)/2
=\nu$ (resp. $s=2t, 1\le t\le (j-7/2)/2=\nu-1$). This allows one to compute $A$ as 
a sum of first, second, and third powers of terms in arithmetic progression, 
which is a simple operation. Namely we get 
\begin{subequations}\begin{align}\label{eq:valA}
 A&=\frac{23j^4}{288}-\frac{23j^3}{144}+\frac{49j^2}{576}-\frac{139j}{576}+\alpha\\
 \alpha&=\frac{2063}{4608}\text{ if }j-1/2\text{ even},\quad
 \frac{1919}{4608}\text{ if }j-1/2\text{ odd.}
\end{align}\end{subequations}
For $j\ge7/2$, specifying the contributions $u=0,1$ the sum $\Xi$ may be written 
$\Xi=\Xi_1+\Xi_0, \Xi_1=\sum_{t=1}^{j-5/2}\xi(t), \Xi_0=\sum_{t=2}^{j-3/2}\xi(t)$ .
The quantities $\Xi_0,\Xi_1,\Xi$ as functions of $j$ are easily derived from the 
definition (\ref{eq:xi}) of $\xi$. Since $\xi(n)$ is periodic with period 6, one
will note that $\sum_{n=0}^5\xi(n)=-3/8$, and therefore $\Xi_1(j+6)=\Xi_1(j)-3/8$, 
$\Xi_0(j+6)=\Xi_0(j)-3/8$, and $\Xi(j+6)=\Xi(j)-3/4$.

We have, from the definition (\ref{eq:defXi}), defining a new table $T_\xi(n)$ 
equally periodic with period 6,
\begin{multline}\label{eq:valxi}
 \Xi=\sum_{t=0}^{j-7/2}\left[\xi(2+t)+\xi(1+t)\right]
 =-\frac34\left\lfloor\frac{j-7/2}{6}\right\rfloor+T_\xi(j-7/2)
 =-\frac18\left[j-\frac72-\left(j-\frac72\bmod6\right)\right]+T_\xi(j-7/2)\\
 \text{where }T_\xi(n)=\left(-\frac{1}{72},-\frac{1}{4},-\frac{3}{8},
 -\frac{7}{18}-\frac{5}{8},-\frac{3}{4}\right)
 \text{ if }n\bmod6=(0,1,2,3,4,5)\text{ respectively.}
\end{multline}
The sum over $\omega(n)$ is obtained in a similar way. One has, from 
definition (\ref{eq:defOmega}),
\begin{equation}
 \Omega=\sum_{s=1}^{j-5/2}\left[\omega(3j-3s-5/2)+\omega(3j-3s-3/2)\right]
 =\sum_{t=0}^{j-7/2}\left[\omega(5+3t)+\omega(6+3t)\right]=\Omega_1+\Omega_0
\end{equation}
and using the $\omega$-value (\ref{eq:omega}) it easy to check that $\Omega_1(j+4)=
\Omega_1(j)+4/9$, $\Omega_0(j+4)=\Omega_0(j)$, $\Omega(j+4)=\Omega(j)+4/9$.  
One obtains the last contribution 
\begin{equation}
 \Omega=\frac49\left\lfloor\frac{j-7/2}{4}\right\rfloor+U_\omega(j-7/2)
 \text{ with }U_\omega(n)=\left(\frac{17}{72},\frac{17}{36},\frac{11}{24},
 \frac{4}{9}\right)
 \text{ if }n\bmod4=(0,1,2,3)\text{ respectively.}
\end{equation} 
With $\lfloor(j-7/2)/4\rfloor=(j-7/2-(j-7/2\bmod4))/4$, one gets 
\begin{equation}\label{eq:valomega}
\Omega=\frac19\left(j-\frac72-\left(j-\frac72\bmod4\right)\right)+U_\omega(j-7/2).
\end{equation}
Collecting (\ref{eq:valomega}), (\ref{eq:valxi}), one obtains
\begin{subequations}\label{eq:OmegaXi}\begin{equation}
\Omega-\Xi=\frac{17}{72}j
 -\frac{119}{144}+\mathscr{T}(j-7/2\bmod6)+\mathscr{U}(j-7/2\bmod4)
\end{equation}
with, for $n=0,1,2,3,4,5$, $\mathscr{T}(n)=-T_\xi(n)-n/8$ or
\begin{equation}
\mathscr{T}(n)=
\left(\frac{1}{72},\frac18,\frac18,\frac{1}{72},\frac18,\frac18\right),
\end{equation}
and, for $n=0,1,2,3$, $\mathscr{U}(n)=U_\omega(n)-n/9$ or
\begin{equation}
\mathscr{U}(n)=
\left(\frac{17}{72},\frac{17}{36},\frac{17}{72},\frac19\right).
\end{equation}\end{subequations}
The expression for $P(1/2;j,5)=A-\Xi+\Omega$ comes from relations 
(\ref{eq:valA}), (\ref{eq:OmegaXi}). We get
\begin{subequations}\begin{equation}\label{qtotj5}
 P(1/2;j,5)=
  \frac{23j^4}{288}-\frac{23j^3}{144}+\frac{49j^2}{576}-\frac{j}{192}+p_0
\end{equation}
with
\begin{multline}
p_0=\left(-\frac{73}{512},-\frac{737}{4608},\frac{55}{512},-\frac{25}{512},
 -\frac{1169}{4608},-\frac{25}{512},\frac{55}{512},-\frac{737}{4608},
 -\frac{73}{512},-\frac{25}{512},-\frac{17}{4608},-\frac{25}{512}\right)\\
 \text{if }j-\frac52\bmod12=(0,1,2,3,4,5,6,7,8,9,10,11).
\end{multline}\end{subequations}
For instance, one gets $P(1/2;7/2,5)=6$.
Since a $j=7/2$ subshell has a degeneracy $g=8$, this corresponds to a three-hole 
system. One expects that the total number of levels is the same for a three-fermion $j=7/2$ shell. 
Using Eq. (\ref{qtotj3}) one indeed finds $P(1/2;7/2,3)=6$, in agreement with the $j^5$ number 
of levels. This is a simple consistency check of Eq. (\ref{qtotj5}).

\section{Derivation of sum rules for six-\textit{j} and nine-\textit{j} 
symbols}\label{sec:sum}

\subsection{Three-fermion case: Sum rules for six-$j$ symbols}

It was shown in Ref. \cite{Pain2019} that, for three-fermion systems,
\begin{equation}\label{qjj3}
Q(J,j,3)=\frac{1}{3}
\sum_{\substack{J_\text{min}\le J_1\le J_\text{max}\\J_1\text{ even}}}
\left[1+2(2J_1+1)\sixj{J_1}{j}{J}{J_1}{j}{j}\right],
\end{equation}
where $J_{\text{min}}=\left|J-j\right|$ and $J_{\text{max}}=\min(2j,j+J)$. Replacing the left-hand side of Eq. (\ref{qjj3}) by the expressions (\ref{eq:QjpqjN3}) and (\ref{eq:QjmqjN3}) of $Q(J,j,3)$ provides a new sum rule on six-$j$ coefficients
\begin{equation}
2\sum_{\substack{J_\text{min}\le J_1\le J_\text{max}\\J_1\text{ even}}}
\left[1+2(2J_1+1)\sixj{J_1}{j}{J}{J_1}{j}{j}\right]=
\begin{cases}
(3j-J)+\tilde{q}_{3p}\text{ with } \tilde{q}_{3p}=\left(0,-1,-2,3,-4,1\right)\text{ for }j\ge J\\
\text{ and }(3j-J)\bmod6=(0,1,2,3,4,5)\text{ respectively},\\
\\
2J+\tilde{q}_{3m}\text{ with } \tilde{q}_{3m}=\left(-1,3,1\right)\text{ for }\text{ for }j\le J\\
\text{ and }J\bmod3=(1/2,3/2,5/2)\text{ respectively}.
\end{cases}
\end{equation}

%
To our knowledge, the above sum rule is not included in reference books such as Ref. \cite{Varshalovich1988}, nor can be deduced in a simple way from elementary sum rules. 
\subsection{Four-fermion case: Connection to Ginocchio-Haxton and 
Rosensteel-Rowe sum rules}\label{gxrr}

The number of $J$=0 states for four fermions in a single-$j$ shell was originally solved by Ginocchio and Haxton \cite{Ginocchio1993, Zamick2005, Pain2018}. They found that
\begin{equation}\label{s1}
Q(0;j,4)=\left\lfloor\frac{2j+3}{6}\right\rfloor.
\end{equation}
Using formula (\ref{eq:Q2jmnjN4}) with $n=2j$, one gets after simple operations $Q(0;j,4)=j/3-1/12+\overline{\omega}(4j-1)-\overline{\xi}(2j)$. With the above definitions of $\overline{\xi}$ and $\overline{\omega}$, one gets $Q(0;j,4)=(j-1/2)/3$, $(j+3/2)/3$ and $(j+1/2)/3$ for $j-1/2\bmod3=0,1,2$ respectively. It is then simple to verify that such expressions are identical to $\lfloor j/3+1/2\rfloor$.
Rosensteel and Rowe showed that the number of linear constraints and algebraic expressions for conservation of seniority can be derived with the quasi-spin tensor decomposition of the two-body interaction. They proposed a matrix which can project the eigenvectors to two quasi-spin subspaces, stated that the eigenvalues of the matrix must equal to 2 or $-1$ and showed that way \cite{Rosensteel2003} that the number of $J$=0 states for four fermions is equal to
\begin{equation}\label{s2}
Q(0;j,4)=\frac{1}{3}\left(\frac{2j+1}{3}
 +2\sum_{\text{even } J_0}(2J_0+1)\sixj{j}{j}{J_0}{j}{j}{J_0}\right).
\end{equation}
From Eqs. (\ref{s1}) and (\ref{s2}), Zhao pointed out that
\begin{equation}
\sum_{\text{even }J_0}(2J_0+1)\sixj{j}{j}{J_0}{j}{j}{J_0}
\end{equation}
has a modular behavior \cite{Zhao2003b} (the sum over all $J_0$ was calculated by 
Schwinger for instance \cite{Schwinger1965} but none of these sums --- over all 
values of $J_0$ or over even values only --- are given in the handbook by 
Varshalovich \textit{et al} \cite{Varshalovich1988}). The values are $(-0.5,0.5,0)$ 
for $j$ values $(1/2,3/2,5/2)$, and repeat after that, i.e., are the same for $j$ 
values $(7/2,9/2,11/2)$, $(13/2,15/2,17/2)$, $(19/2,21/2,23/2)$, etc. The first 
three values $-0.5$, $0.5$, $0$ for $j=1/2$, $3/2$ and $5/2$ respectively were 
obtained by Zamick and Escuderos using recursion relations for coefficients of 
fractional parentage \cite{Zamick2005c, Zamick2006}.
Noticing that the number of $J=j$ states for three fermions in equal to the number of $J=0$ states for four fermions, Zamick and Escuderos proposed an alternate derivation \cite{Zamick2005} of $Q(0;j,4)$. 
In 2010, Qi \emph{et al} published an alternative proof of the Rosensteel-Rowe relation relying on a decomposition of the total angular momentum. In this work, a matrix similar to that of Ref.~\cite{Zhao2003} has been constructed from the decomposition and the eigenvalue problem was explored in a general way with symmetry properties of angular-momentum coupling coefficients \cite{Qi2010}.
All those properties (Ginocchio-Haxton and Rosensteel-Rowe relations, sum rules over six-$j$ symbols) are obtained in a straightforward way by the formulas given in the preceding sections.

\subsection{Four-fermion case: Sum rules for nine-\textit{j} symbols}
In the same paper \cite{Pain2019}, the following expression was derived for $j^4$
\begin{equation}
Q(J,j,4)=\frac{1}{6}\sum_{\substack{J_1\text{ even}\\0\le J_1\le2j}}
\sum_{\substack{J_2\text{ even}\\0\le J_2\le2j}}\Delta(J_1,J_2,J)
\left[1+(-1)^{J}\delta_{J_1,J_2}-4(2J_1+1)(2J_2+1)
\ninej{j}{j}{J_2}{j}{j}{J_1}{J_2}{J_1}{J}\right]
\end{equation}
where $\Delta(J_1,J_2,J)=1$ if $(J_1,J_2,J)$ verify the triangular conditions, 
0 otherwise.
Setting $J=2j-n$ in Eq.(\ref{eq:Q2jmnjN4}), we get the sum rule
\begin{align}\label{sr40}
&4\sum_{\substack{J_1\text{ even}\\0\le J_1\le2j}}
\sum_{\substack{J_2\text{ even}\\0\le J_2\le2j}}
\Delta(J_1,J_2,J)
\left[1+(-1)^{J}\delta_{J_1,J_2}
 -4(2J_1+1)(2J_2+1)\ninej{j}{j}{J_2}{j}{j}{J_1}{J_2}{J_1}{J}\right]\nonumber\\
& =2\left(2j-\frac{J+3}2\right)^2-\frac{7}{3}
 -H(2j-J)\left[2(2j-J-1)^2-\frac{7}{3}+24\overline{\xi}(2j-J)\right]\nonumber\\
&\quad +24\overline{\omega}(4j-J-1) \begin{cases}
 &\text{ if } (2j-J)\text{ even},\\
+3(4j-J-3/2)&\text{ if } (2j-J)\text{ odd}.\end{cases}
\end{align}
As implied by the triangular and parity conditions, the above relation is derived 
assuming that $J\le4j-2$. For higher $J$, the left-hand side always vanishes while 
the right-hand side does vanish if $4j-1\le J\le4j+2$, but equals 24 if $J=4j+3$.   
The total number of levels in $j^4$ reads
\begin{subequations}\label{sumj4}\begin{align}
Q_{\mathrm{tot}}(j^4)&=\sum_{J=0}^{2(2j-3)}Q(J,j,4)\\
 &=
 \frac{1}{72}(2j+1)\left(8j^2+2j+9\right)
  -\frac{2}{3}\sum_{J_1,J_2\text{ even}} (2J_1+1)(2J_2+1)
\sum_{J=|J_1-J_2|}^{J_1+J_2}\ninej{j}{j}{J_2}{j}{j}{J_1}{J_2}{J_1}{J},
\end{align}\end{subequations}
and therefore expression (\ref{qtotj4}) enables one to write the sum rule
\begin{equation}\label{sr41}
4\sum_{J_1,J_2\text{ even}}
(2J_1+1)(2J_2+1)\sum_{J=|J_1-J_2|}^{J_1+J_2}\ninej{j}{j}{J_2}{j}{j}{J_1}{J_2}{J_1}{J}
=2j^2+\frac{2}{3}j
 \begin{cases}+\frac{7}{6}&\quad\text{ if }\left(j-\frac12\right)\bmod3=0,\\
 -\frac{3}{2}&\quad\text{ if }\left(j-\frac12\right)\bmod3=1,\\
 -\frac{1}{6}&\quad\text{ if }\left(j-\frac12\right)\bmod3=2.\end{cases}
\end{equation} 
Equation (\ref{sumj4}) can also be expressed using the coefficients introduced by Dunlap and Judd \cite{Dunlap1975}
\begin{equation}\label{dun}
D_{J_a,J_b;k}=\frac{1}{2k+1}\left[\frac{\left(2J_a-k\right)!\left(2J_b+k+1\right)!}{\left(2J_b-k\right)!\left(2J_a+k+1\right)!}\right]^{1/2},
\end{equation}
as
\begin{align}\label{sumdun}
Q_{\mathrm{tot}}\left(j^4\right)&=\frac{2j+1}{72}\left[2j(4j+1)+9\right]
\nonumber\\
&-\frac23\sum_{\substack{J_1,J_2\\J_1,J_2\text{ even}}}
  (2J_1+1)(2J_2+1)\sum_{k=0}^{\min(2j,2J_1,2J_2)}(2k+1)(-1)^{\phi}D_{J_M,J_m;k}\sixj{j}{j}{k}{J_2}{J_2}{j}\sixj{j}{j}{k}{J_1}{J_1}{j}
\end{align}
with $\phi=J_1+J_2+k$, $J_m=\min\left(J_1,J_2\right)$, $J_M=\max(J_1,J_2)$. 
Of course the sum is restricted to conditions $0\le J_1\le2j$, $0\le J_2\le2j$ 
imposed by the 6-$j$ symbol. The corresponding sum rule is therefore
\begin{align}\label{sr42}
&4\sum_{\substack{J_1,J_2\\J_1,J_2\text{ even}}}
(2J_1+1)(2J_2+1)
\sum_{k=0}^{\min(2j,2J_1,2J_2)}(2k+1)(-1)^{\phi}D_{J_M,J_m;k}\sixj{j}{j}{k}{J_2}{J_2}{j}\sixj{j}{j}{k}{J_1}{J_1}{j}\nonumber\\
&=2j^2+\frac{2}{3}j
 \begin{cases}+\frac{7}{6}&\text{ if }\left(j-\frac12\right)\bmod3=0,\\
 -\frac{3}{2}&\text{ if }\left(j-\frac12\right)\bmod3=1,\\
 -\frac{1}{6}&\text{ if }\left(j-\frac12\right)\bmod3=2.\end{cases}
\end{align}
To our knowledge, Eqs. (\ref{sr40}), (\ref{sr41}) and (\ref{sr42}) are not included in reference books such as Ref. \cite{Varshalovich1988}, nor can they 
be deduced in a simple way from elementary sum rules.

\section{Particular values of the number of levels with a given spin \textit{J}}
\label{sec:part}

A property mentioned by Talmi is the vanishing of $Q(1/2;j,3)$. It is worth mentioning that for $j=1/2$, it is not possible to get three distinct values $m_1,m_2,m_3$ because $m_i=\pm1/2$ and therefore $Q(1/2;1/2,3)=0$. From the above relation (\ref{eq:QvsP}), one also gets 
\begin{equation}\label{eq:TalmiQ}
 Q(J;j,N)=Q(J;j-1,N)+Q(J-j;j-1,N-1)+Q(J+j;j-1,N-1)+Q(J;j-1,N-2).
\end{equation}
The recurrence (\ref{eq:TalmiQ}) reads, for $J=1/2, N=3$ and accounting for the formal symmetry property $Q(-J-1)=-Q(J)$, 
\begin{subequations}\begin{align}\label{eq:recQj_jm1N3}
 Q\left(\frac12;j,3\right)&=Q\left(\frac12;j-1,3\right)+Q\left(\frac12-j;j-1,2\right)+Q\left(\frac12+j;j-1,2\right)+Q\left(\frac12;j-1,1\right)\\
  &=Q\left(\frac12;j-1,3\right)-Q\left(j-\frac32;j-1,2\right)+Q\left(j+\frac12;j-1,2\right)+Q\left(\frac12;j-1,1\right).\label{eq:Qj1s2}
\end{align}\end{subequations}
Let us note first that the fourth term of that equation is zero except if $j=3/2$. For $j=3/2$, the second term is $-Q(0;1/2,2)=-1$, the third $Q(2;1/2,2)=0$, and the fourth $Q(1/2;1/2,1)=1$ according to the elementary properties of the coupling of angular momenta $j=1/2$. The sum  of the last three terms of (\ref{eq:Qj1s2}) is therefore zero. For $j=5/2$, $-Q(1;3/2,2)=Q(3;3/2,2)=0$ because the total momentum $J$ must be even, and the sum of the last three terms of (\ref{eq:Qj1s2}) cancels as well. For $j\ge7/2$, one has $j-3/2<j+1/2\le J_\text{max}=2j-3$. For $J\ge0$, 
$Q(J;j-1,2)=1$ if $J$ even, 0 otherwise. One checks 
\begin{equation}
-Q\left(j-\frac32;j-1,2\right)+Q\left(\frac12+j;j-1,2\right)=\begin{cases}
-1+1=0&\text{ for }j=2n-\frac12\\
 0+0=0&\text{ for }j=2n+\frac12\end{cases}
\end{equation}
and therefore for each $j$ the summation of the last three terms of (\ref{eq:Qj1s2}) cancels. Such an equation implies that for $j$ half-integer
\begin{equation}
Q\left(\frac12;j,3\right)=Q\left(\frac12;j-1,3\right)=\cdots=Q\left(\frac12;\frac32,3\right)=Q\left(\frac12;\frac12,3\right)=0.
\end{equation}
%
In addition, it is easy to show that $Q(J_\text{max}(j,N);j,N) = 1$ and that 
  $Q(J_\text{max}(j,N)-1;j,N) = 0$.
Indeed, for each configuration 
$j^N$, the value $J=J_\text{max}$ is realized only once. This manifests clearly if 
one notes that in order to get $M=J_\text{max}$ there is only one solution except 
permutations of the $m_i$, which is $m_1=j-N+1, m_2=j-N+2, \cdots, m_N=j$, yielding 
$P\left(J_\text{max};j,N\right)=1$. For $M=J_{\mathrm{max}}-1$, the only possibility 
is to reduce $m_1$ by one with respect to the $J_{\mathrm{max}}$ case: $m_1=j-N, 
m_2=j-N+2, \cdots, m_N=j$ and one has also $P(J_\text{max}-1;j,N)=1$ and thus 
$Q(J_\text{max}-1;j,N)=0$.

\section{Conclusion}
Closed-form expressions for the number of levels for three, four and five fermions 
in a single-$j$ shell are obtained using recursion relations for $P(M)$, the number 
of states with a given magnetic quantum number $M$. We derive exact expressions 
for $P(M)$ and $Q(J)$, the number of levels with a given total angular momentum $J$, 
in the cases of $j^3$ and $j^4$. The formulas involve polynomials, the coefficients 
of which are defined by congruence relations. We provide supplementary results, 
such as proofs of empirical formulas published by several authors over the last 
years, cancellation properties and peculiar values of $Q(J)$, or new sum rules over 
six-$j$ and nine-$j$ symbols.

\appendix
\section{Recurrence relation on the number of fermions for the quantum numbers \textit{j} and \textit{j}--1}\label{sec:Talmirec}
We have established in Appendix B of Ref.~\cite{Poirier2021a} 
the two relations (respectively (B4) and (B8))
\begin{subequations}\begin{align}
 P(M;j,N)&=P\left(M-\frac{N}{2};j-\frac12,N\right)
  +P\left(M-\frac{N}{2}+j+\frac12;j-\frac12,N-1\right),\label{eq:B4}\\
P(M;j,N)&=P\left(M+\frac{N}{2};j-\frac12,N\right)
 +P\left(M+\frac{N}{2}-j-\frac12;j-\frac12,N-1\right)\label{eq:B8}
\end{align}\end{subequations}
from the recurrences for the Gaussian binomial coefficient. 
The first term on the right-hand-side of (\ref{eq:B4}) can be transformed using (\ref{eq:B8})
\begin{equation}
P\left(M-\frac{N}{2};j-\frac12,N\right)=P(M;j-1,N)+P(M-j;j-1,N-1).\label{eq:recj_t1}
\end{equation}
In the same way, the second term on the right-hand-side of (\ref{eq:B4}) transforms 
with (\ref{eq:B8}) into
\begin{subequations}
\begin{align}\nonumber
P\left(M-\frac{N}{2}+j+\frac12;j-\frac12,N-1\right)&=
P\left(M-\frac{N}{2}+j+\frac12+\frac{N-1}{2};j-1,N-1\right)\\
&\quad+P\left(M-\frac{N}{2}+j+\frac12+\frac{N-1}{2}-j;j-1,N-2\right)\\
&=P(M+j;j-1,N-1)+P(M;j-1,N-2)\label{eq:recj_t2}
\end{align}
\end{subequations}
and gathering equations (\ref{eq:B4}), (\ref{eq:recj_t1}), (\ref{eq:recj_t2}), 
we get the basic equation (\ref{eq:Talmi})
which was previously obtained by Talmi [Eq.~(1) of Ref.~\cite{Talmi2005}].

\section{Examples of $P(M;j,3)$ values for $M\le11/2$}\label{sec:PlowMj3}
The relation (\ref{eq:PjmqjN3}) may be used to get $P(j-q;j,3)$ for $j-q=1/2, 
3/2,\dots n+1/2$. Examples for the first $j-q$ values are given in 
Table \ref{tab:PMj3}, with the notation $P(M;j,3)=\frac12(j^2-c_M/4)$, and 
assuming $M\le j$. For instance $P(11/2;j,3)=\frac12(j^2-41/4)$ only if $j\ge11/2$. 
One calculates $P(11/2;7/2,3)=2$, although this formula would give 1.
We obtain again from $P(1/2;j,3)$ the total number of levels for three 
fermions derived above (\ref{qtotj3}) and also obtained in Ref.~\cite{Pain2019} 
using fractional parentage coefficients. 
\begin{table}[htb]
\centering
\begin{tabular}{c@{\quad}cccccc}
\hline\hline
$M$ & 1/2 & 3/2 & 5/2 & 7/2 & 9/2 & 11/2 \\
$c_M$ & 1 &   1 &   9 &  17 &  25 &  41\\
\hline\hline
\end{tabular}
\caption{Coefficient for the three-fermion distribution of the quantum magnetic 
number for the lowest $M$ values. This number is given by $P(M;j,3)=\frac12(j^2-c_M/4)$, assuming $j\ge M$.\label{tab:PMj3}}
\end{table}

\section{Determination of the distribution $P(2j-1;j,4)$}
\label{sec:P2jm1j4}
The value $P(2j-1;j,4)$ is derived starting from Eq.~(\ref{eq:P2jpj4o}) 
that can be rewritten as 
\begin{subequations}\label{eq:P2jp1j4}\begin{gather}
P(2j+1;j,4)=\frac{(j-1)^3}{18}-\frac{(j-1)}{24}+\varphi,\\
\varphi=\left(-\frac{1}{72},\frac{1}{72},-\frac18,\frac{17}{72},
-\frac{17}{72},\frac18\right)
\text{ if }j-\frac12\bmod6=(0,1,2,3,4,5)\text{ respectively}.
\end{gather}\end{subequations}
We will obtain $P(2j-1;j,4)$ from the fundamental equation (\ref{eq:Talmi}), 
and the definition (\ref{eq:defF})
\begin{equation}P(2j-1;j,4)=P(2j-1;j-1,4)+P(j-1;j-1,3)+F(j,1).\end{equation}
We note that $F(j,1)=0$ according to (\ref{eq:Fj2q1}). With the notations $x=j-2$, 
$P_1=P(2j-1;j-1,4)$, $P_3=P(j-1;j-1,3)$, and the value (\ref{eq:P2jp1j4}) 
for $P_1$ we obtain the following results.
\begin{itemize}
\item[$\bullet$]If $j\bmod6=1/2$, 
 $P_1=x^3/18-x/24+1/8, P_3=(j-1)^2/3-1/12$, 
 $P(2j-1;j,4)=j^3/18-j/24+1/72$,
\item[$\bullet$]if $j\bmod6=3/2$, 
$P_1=x^3/18-x/24-1/72, P_3=(j-1)^2/3-1/12$,
$P(2j-1;j,4)=j^3/18-j/24-1/8$,
\item[$\bullet$]if $j\bmod6=5/2$, 
$P_1=x^3/18-x/24+1/72, P_3=(j-1)^2/3+1/4$,
$P(2j-1;j,4)=j^3/18-j/24+17/72$,
\item[$\bullet$]if $j\bmod6=7/2$,
$P_1=x^3/18-x/24-1/8, P_3=(j-1)^2/3-1/12$,
$P(2j-1;j,4)=j^3/18-j/24-17/72$,
\item[$\bullet$]if $j\bmod6=9/2$,
$P_1=x^3/18-x/24+17/72, P_3=(j-1)^2/3-1/12$, 
$P(2j-1;j,4)=j^3/18-j/24+1/8$,
\item[$\bullet$]if $j\bmod6=11/2$,
$P_1=x^3/18-x/24-17/72, P_3=(j-1)^2/3+1/4$, 
$P(2j-1;j,4)=j^3/18-j/24-1/72$.\end{itemize}
These expressions are needed for initializing the recurrence (\ref{eq:hypP2jm2q1}).

%

\end{document}